\documentclass[journal, onecolumn, 12pt, letterpaper]{IEEEtran}
\usepackage{amsmath,amsfonts}
\usepackage{algorithmic}
\usepackage{algorithm}
\usepackage{array}
\usepackage[caption=false,font=normalsize,labelfont=sf,textfont=sf]{subfig}
\usepackage{textcomp}
\usepackage{stfloats}
\usepackage{url}
\usepackage{verbatim}
\usepackage{graphicx}
\usepackage{cite}
\usepackage{amssymb}
\usepackage{bbm}
\usepackage{makecell}
\usepackage{multirow, multicol}
\usepackage{pgf}
\usepackage{xcolor, colortbl}

\newtheorem{theorem}{Theorem}[section]

\newtheorem{lemma}[theorem]{Lemma}
\newtheorem{proposition}[theorem]{Proposition}
\newtheorem{remark}{Remark}
\newtheorem{definition}{Definition}
\newtheorem{example}{Example}
\newtheorem{conjecture}{Conjecture}

\newcommand{\expect}{\mathbb{E}}

\DeclareMathOperator*{\argmin}{arg\,min}

\begin{document}

\title{Exactly Optimal and Communication-Efficient \\Private Estimation via Block Designs}

\author{
Hyun-Young~Park,~\IEEEmembership{Graduate Student Member,~IEEE,}
Seung-Hyun~Nam,~\IEEEmembership{Graduate Student Member,~IEEE,}
and Si-Hyeon~Lee,~\IEEEmembership{Senior Member,~IEEE}
\thanks{H.-Y. Park, S.-H. Nam, and S.-H. Lee are with the School of Electrical Engineering, Korea Advanced Institute of Science and Technology (KAIST), Daejeon 34141, South Korea  (e-mail: phy811@kaist.ac.kr; shnam@kaist.ac.kr; sihyeon@kaist.ac.kr). (Corresponding Author: Si-Hyeon Lee)}
\thanks{A short version of this manuscript was presented at 2023 IEEE International Symposium on Information Theory \cite{THIS-PAPER-ISIT-VER}. {The added contents over the conference paper include not only the full proofs of the theorems and propositions, but also the following new results: (1) a way of obtaining a non-symmetric block design achieving exact or near-optimality by considering derived and residual block designs in Section~\ref{subsec:symmetric}, (2) a discussion on the Hadamard conjecture in Section \ref{subsec:hadConj}, (3)  a characterization of the exact optimality condition in Proposition \ref{prop:optEpsRegion}, and (4) efficient algorithms with the analysis of computational complexities for the proposed schemes in Appendix \ref{app:blockDesignsAlgorithms} of the supplementary material.}}
\thanks{This article has supplementary material provided by the authors.}%
}

\maketitle

\begin{abstract}
In this paper, we propose a new class of local differential privacy (LDP) schemes based on combinatorial block designs for  discrete distribution estimation.
This class not only recovers many known LDP schemes in a unified framework of combinatorial block design, but also suggests a novel way of finding new schemes achieving the {exactly} optimal (or near-optimal) privacy-utility trade-off with lower communication costs.
Indeed, we find many new LDP schemes that achieve {the {exactly} optimal privacy-utility trade-off, with the minimum communication cost among all the unbiased {or consistent} schemes,} for a certain set of input data size and LDP constraint.
Furthermore, to partially solve the sparse existence issue of block design schemes, we consider a broader class of LDP schemes based on regular and pairwise-balanced designs, called RPBD schemes, which relax one of the symmetry requirements on block designs. By considering this broader class of RPBD schemes, we can find LDP schemes achieving near-optimal privacy-utility trade-off with reasonably low communication costs for a much larger set of input data size and LDP constraint.
\end{abstract}


\section{Introduction}
\IEEEPARstart{S}{tatistical} inference often involves the collection of private data about individuals, e.g., health conditions, preference on merchandises, search histories, etc.
It has been continuously reported that a single data in a large dataset can be inferred from the anonymized dataset, statistics about the dataset, and even  machine learning models trained with the dataset \cite{NetflixPrizeBreakAnonymity-06Narayanan, modelInversionConfInfo-15Fredrikson, MLMembershipInfer-17Shokri, MLPrivacyOverfit-18Samuel, updatesLeak-20Salem, CollLearningFeatureLeakage-19Melis, invertingGradients-20Geiping}, which necessitates the application of appropriate privacy protection techniques.
An effective and powerful privacy protection technique is to perturb each data to satisfy a certain privacy constraint before sent to the data collector. 
Among many notions of privacy constraints \cite{alpha_leakage-19Liao, InfoLeakOperational-20Issa, LaplaceLDP-06Dwork, RenyiDP-17Mironov}, local differential privacy (LDP) is widely adopted and applied both in academia and industry due to its powerful privacy guarantee, operational meaning, and useful properties such as composability and robustness to post-processing \cite{InfoLeakOperational-20Issa, compositionDP-13Sewoong}.
An LDP mechanism is applied at each client side, requiring that each possible true data (input of the mechanism) should induce a similar distribution on the data sent to the collector (output of the mechanism). 
The LDP constraint was shown to upper-bound the probability of correctly inferring any information from the protected data \cite[Theorem 14]{InfoLeakOperational-20Issa}.
As such, LDP guarantees a fundamental privacy protection against the data collector and any other attackers. 
Furthermore, LDP is well-suited for statistical inference systems because the definition of LDP does not depend on the underlying distribution of data, while some other privacy constraints based on mutual information and $\alpha$-leakage \cite{alpha_leakage-19Liao} depend on the prior distribution.

Many LDP mechanisms and corresponding estimators have been proposed for various statistical inference tasks, such as discrete distribution estimation\cite{RR-65Stanley, RAPPOR-14Erlingsson, SS-18MinYe, pureMech-17Wang, HR-19Acharya, RHR-20WeiNingChen, PGR-22Feldman},  mean estimation\cite{LaplaceLDP-06Dwork, DuchiMech-13Duchi, piecewiseMech-19Wang, PrivUnit-18Bhowmick, RHR-20WeiNingChen}, etc. Let us call a pair of an LDP mechanism at the clients and the  estimator at the data collector as an (LDP) scheme. 
Note that a stronger privacy constraint requires more perturbation of the data, which in turn causes more performance degradation in statistical inference, i.e., more loss of utility. Such a fundamental trade-off between privacy and utility has been actively studied 
\cite{LDPStatMinimax-13Duchi, FIunderLDP-20Barnes, SS-18MinYe, SSOptim-19MinYe, PrivUnit-18Bhowmick, PrivUnitOptim-22Asi}. 
In particular, the subset selection (SS) scheme proposed in \cite{SS-18MinYe} was shown to achieve the exactly optimal privacy-utility trade-off for the discrete distribution estimation \cite{SSOptim-19MinYe}. 
For the mean estimation on spherical data, PrivUnit$_2$ introduced in \cite{PrivUnit-18Bhowmick} was proved to be exactly optimal among all the unbiased schemes  \cite{PrivUnitOptim-22Asi}.

In addition to privacy and utility, another factor of practical importance is the communication cost for sending data from each client to the data collector \cite{CommEffDistMLParamServer-14LiMu, CommEffDistStatInfer-19Michael, CommEffStatOpt-12Zhang}.  
The {exactly} optimal LDP schemes with respect to privacy-utility trade-off, however, are known to require very large communication costs, i.e., the SS \cite{SS-18MinYe} requires the communication cost of order greater than the exponential of the input data size and PrivUnit$_2$ \cite{PrivUnit-18Bhowmick} produces continuous outputs. 
Recently, many communication-efficient LDP schemes have been proposed for private statistical inference \cite{HR-19Acharya, RHR-20WeiNingChen, PGR-22Feldman, one_bit_LDP-15Bassily, MMRC-22Shah}.
Furthermore, these schemes were shown to achieve the order-optimal privacy-utility trade-off, i.e., achieve the optimal trade-off up to some multiplicative constant factors.
{Let us provide a more detailed description of related works in Section \ref{subsec:prevWorks}, after we formally define the privacy-utility trade-off in Section  \ref{subsec:systemModel}.} In practice, however, analyzing the exact privacy-utility trade-off including the leading constant factor, not in the order-optimal sense, is important as the effect of constant factors becomes non-negligible. 

In this paper, we propose a new class of LDP schemes based on combinatorial block designs \cite{CombSymmDesigns-06Ionin} for discrete distribution estimation, which we call block design schemes. This class not only recovers many known LDP schemes in a unified framework of combinatorial block design, including randomized response (RR)\cite{RR-65Stanley}, SS\cite{SS-18MinYe}, Hadamard response (HR)\cite{HR-19Acharya}, projective geometry response (PGR) \cite{PGR-22Feldman}, to name a few, but also suggests a novel way of finding new schemes achieving the {exactly} optimal (or near-optimal) privacy-utility trade-off,  with lower communication costs. 
The proposed  block design scheme has six parameters, i.e., sizes of the input and the output of the mechanism, LDP constraint, and three parameters related to some symmetry properties. Since the input data size (the input of the mechanism) and the LDP constraint are predetermined, we have four design parameters to choose. 
We show that the risk is determined by only a single parameter among the four, i.e.,  one of the symmetry-related parameters, called uniformity parameter.
This fact opens up possibilities of finding schemes achieving {exactly} optimal (or near-optimal) privacy-utility trade-off with lower communication costs by considering block designs with the same (or similar) uniformity parameter with the optimal SS \cite{SS-18MinYe}.
Indeed, we find many new LDP schemes in this class that {are exactly optimal, with the minimum communication cost  among all consistent or unbiased schemes (characterized by \cite{comm_unbiased-19Acharya})}, for a larger set of input data size and LDP constraint than \cite{PGR-22Feldman}.

In general, however, due to the necessary conditions that the parameters of a block design should satisfy, block designs exist for a sparse set of parameters. Thus, a block design scheme achieving {the exactly optimal privacy-utility trade-off} with reasonably low communication cost may not exist for some input data size and LDP constraint. To overcome this issue of sparse existence of block designs, we consider regular and pairwise-balanced design (RPBD) schemes, which relaxes one of the symmetry requirements on block designs. By doing so, we indeed find new privacy schemes achieving near-optimal privacy-utility trade-off with reasonably low communication costs for a much larger set of input data size and LDP constraint.
For example, we find schemes which achieve smaller estimation error while requiring less communication cost compared to previously known schemes (PGR\cite{PGR-22Feldman}) in some medium privacy regime by exploring RPBD schemes.

The rest of this paper is organized as follows. In Section~\ref{sec:problem-formulation}, we formulate the problem of private discrete distribution estimation {and present some related works}.
In Section~\ref{sec:mainSection}, we first introduce some basics of block design, and then present block design schemes and RPBD schemes with sufficient conditions for two schemes to achieve the same risk, which play the key role in finding exactly or nearly optimal schemes with low communication costs.
Finally, with some discussions in Section~\ref{sec:discussion}, Section \ref{sec:conclusion} concludes the paper.

\section{Problem Formulation}\label{sec:problem-formulation}

\subsection{System model}\label{subsec:systemModel}
We consider a discrete distribution estimation problem where a server wants to estimate the underlying data distribution of $n$ clients while preserving each individual's privacy.  
Client $i$ for $i\in [1:n]$ observes $X_i$, which is i.i.d. according to unknown distribution $P \in \Delta_v$ on $\mathcal{X}=[1:v]$, where $v\geq 2$ is finite and
\begin{IEEEeqnarray}{c}
    \Delta_{v} := \left\{P=(P_1,\cdots,P_v) \in \mathbb{R}^v : P_x \geq 0, \sum_{x=1}^{v} P_x = 1\right\}\IEEEeqnarraynumspace
\end{IEEEeqnarray}
is the set of all probability mass functions on $\{1,2,\cdots,v\}$.

Then, it  independently generates a privacy-protected data $Y_i \in \mathcal{Y}$ according to a conditional distribution $Q(\cdot|X_i)$ and sends $Y_i$ to the server, where the set $\mathcal{Y}$ and the conditional distribution $Q:\mathcal{X} \rightarrow \mathcal{Y}$ are known to the server. The privacy requirement is formalized by the \textbf{local differential privacy (LDP)}\cite{LDPStatMinimax-13Duchi}.
\begin{definition}
Let $\epsilon>0$ be given. A conditional distribution $Q:\mathcal{X} \rightarrow \mathcal{Y}$ is said to satisfy the $\epsilon$-local differential privacy (in short, $\epsilon$-LDP) if
\begin{IEEEeqnarray}{c}
    Q(A|x) \leq e^\epsilon Q(A|x'),
\end{IEEEeqnarray}
for all $x,x' \in \mathcal{X}$ and a (measurable) set $A \subset \mathcal{Y}$.
\end{definition}
The server collects $Y_1,Y_2,\cdots,Y_n$ from the clients, and uses them  to construct an estimate $\hat{P}_n(Y_1,\cdots,Y_n)$ of $P$, where $\hat{P}_n$ is the estimator given as a function $\hat{P}_n=(\hat{P}_{n,1},\cdots,\hat{P}_{n,v}):\mathcal{Y}^n \rightarrow \mathbb{R}^v$.
We call a conditional distribution $Q$ and a conditional distribution-estimator pair $(Q,\hat{P}_n)$ as an \textbf{(LDP) mechanism} and an \textbf{(LDP) scheme}, respectively.

Let $\ell:\Delta_{v} \times \mathbb{R}^v \rightarrow \mathbb{R}$ be a (measurable) function called a \textbf{loss function}, which measures the difference between two arguments. The \textbf{risk} of a scheme $(Q,\hat{P}_n)$ under a loss function $\ell$ is defined as
\begin{IEEEeqnarray}{c}
    R_{v, \ell}^{n}(P, Q, \hat{P}_n) := \expect\left[\ell(P, \hat{P}_n(Y_1,\cdots,Y_n))\right],
\end{IEEEeqnarray}
where the expectation is over $X_i \sim P$ and $Y_i \sim Q(\cdot|X_i)$.
Since $P$ is unknown, we consider the \textbf{worst case risk}, which is defined as
\begin{IEEEeqnarray}{c}
    R_{v, \ell}^n(Q, \hat{P}_n) := \sup_{P \in \Delta_{v}} R_{v, \ell}^{n}(P, Q, \hat{P}_n). 
\end{IEEEeqnarray}
In this paper, we mainly focus on the $\ell_u^u$-loss for $1 \leq u \leq 2$,
\begin{IEEEeqnarray}{c} 
        \ell_u^u(p,\hat{p}):=\sum_{x=1}^{v}|p_x-\hat{p}_{x}|^u,
    \end{IEEEeqnarray}
for $p\in \Delta_v$ and $\hat{p}\in \mathbb{R}^v$. {This includes the standard mean squared loss, i.e., $\ell_2^2$, and the total variation distance, which is the half of the $\ell_1$ loss. For simplicity, {we use the notation}}
\begin{align}
    R_{v, u}^{n}(P, Q, \hat{P}_n) &:= R_{v, \ell_u^u}^{n}(P, Q, \hat{P}_n),\\
    R_{v, u}^{n}(Q, \hat{P}_n) &:= R_{v, \ell_u^u}^{n}(Q, \hat{P}_n).
\end{align}

It is shown that the optimal (worst-case) risk under $\epsilon$-LDP constraint for $\ell_u^u$ loss, $1 \leq u \leq 2$  has the order of $\Theta(n^{-u/2})$ \cite{SSOptim-19MinYe}.
Hence,  we say that an LDP scheme $(Q,\hat{P}_n)$ achieves the \textbf{asymptotic (worst-case) risk} given as 
    \begin{align}
        {R}_{v,u} (Q, \hat{P}_n) := \lim_{n \rightarrow \infty} n^{u/2} R_{v, u}^{n}(Q, \hat{P}_n). \label{eq:asympRiskDef}
    \end{align}
Also, we define the \textbf{optimal asymptotic (worst-case) risk} under $\epsilon$-LDP constraint for $\ell_u^u$ loss, $1 \leq u \leq 2$,  
    \begin{align}
        M_{v,u,\epsilon} :=
     \inf_{Q:\epsilon\text{-LDP}} \lim_{n \rightarrow \infty} \inf_{\hat{P}_n}n^{u/2} R_{v, u}^{n}(Q, \hat{P}_n). \label{asymptotic}
    \end{align}
We say that an $\epsilon$-LDP scheme $(Q, \hat{P}_n)$ is \textbf{exactly optimal}  for $\ell_u^u$-loss if ${R}_{v,u} (Q, \hat{P}_n)=M_{v,u,\epsilon}$.

On the other hand, the communication cost of a scheme is measured by the number of possible outputs $|\mathcal{Y}|$ of the mechanism, which we denote by $b$. This is the exponential of the number of required bits {per client} for uncoded transmission, which makes sense because the best possible guess for the distribution of $Y_i$ would be the uniform distribution as $P$ is unknown. We note that it suffices to consider $b<\infty$, since any $\epsilon$-LDP scheme can be approximated with arbitrary precision (in some sense) by another $\epsilon$-LDP scheme with finite $b$\cite[Lemmas 11 and 19]{SS-18MinYe}.

\subsection{Related works}\label{subsec:prevWorks}

{Many LDP schemes have been proposed for discrete distribution estimation, examples of which include randomized response (RR)\cite{RR-65Stanley}, randomized aggregatable privacy-preserving ordinal response (RAPPOR) \cite{RAPPOR-14Erlingsson}, subset selection (SS) \cite{SS-18MinYe}, Hadamard response (HR) and generalized Hadamard response (GHR) \cite{HR-19Acharya}, and projective geometry response (PGR) \cite{PGR-22Feldman}. In addition to proposing achievability schemes, the converse part, giving a lower bound on the optimal risk, has been actively studied \cite{LDPStatMinimax-13Duchi, FIunderLDP-20Barnes, SS-18MinYe, SSOptim-19MinYe}.
The most remarkable result is the work \cite{SSOptim-19MinYe}, which completely characterized $M_{v, u,\epsilon}$ for all $(v,\epsilon)$ and $1 \leq u \leq 2$.}
\begin{theorem}\cite[Theorems 3.2 and 3.3]{SSOptim-19MinYe}\label{thm:SSOptim} For any $v \geq 2$, $\epsilon >0$, and $1 \leq u \leq 2$, we have
\begin{equation}
    M_{v, u,\epsilon} = \min_{k \in \mathbb{Z}, 1 \leq k \leq v-1} v C_u \left(\frac{v-1}{v(e^\epsilon -1)}\right)^u\left(\frac{(ke^\epsilon+v-k)^2}{k(v-k)}\right)^{u/2}, \label{eqn:optM}
\end{equation}
where $C_u := \mathbb{E}|Z|^u$ for the standard Gaussian random variable $Z \sim \mathcal{N}(0,1)$.
\end{theorem}

{An exactly optimal scheme that achieves \eqref{eqn:optM} is the SS scheme \cite{SS-18MinYe}. It was shown in \cite{SSOptim-19MinYe} that the for any $v \geq 2$ and $\epsilon > 0$, the SS scheme $(Q,\hat{P}_n)$ with the subset size parameter $k\in \{1,2,\cdots,v-1\}$ achieves  
\begin{equation}
    {R}_{v,u} (Q, \hat{P}_n) = v C_u \left(\frac{v-1}{v(e^\epsilon -1)}\right)^u\left(\frac{(ke^\epsilon+v-k)^2}{k(v-k)}\right)^{u/2}, ~\forall 1\leq u\leq 2. \label{SSachiev}
\end{equation}
Thus, the SS scheme\cite{SS-18MinYe} with the subset size parameter $k \in K^*_{v,\epsilon}:= \argmin_{k \in \mathbb{Z}, 1 \leq k \leq v-1} \frac{(ke^\epsilon+v-k)^2}{k(v-k)}$ is exactly optimal  for all $1 \leq u \leq 2$.
It was further shown in \cite{SS-18MinYe} that}
\begin{equation}
   K^*_{v,\epsilon} \subseteq {\left\{ \left\lfloor \frac{v}{e^\epsilon+1} \right\rfloor, \left\lceil \frac{v}{e^\epsilon+1} \right\rceil \right\}}. \label{eq:optimUnifParam}
\end{equation}
{For $\ell_2^2$ loss, note that \eqref{eqn:optM} and \eqref{SSachiev} are simplified to }
\begin{equation}
   M_{v, 2,\epsilon}=\frac{(v-1)^2(k^*e^\epsilon+v-k^*)^2}{k^*(v-k^*)(e^\epsilon-1)^2 v},  ~~    {R}_{v,2} (Q, \hat{P}_n) = \frac{(v-1)^2(ke^\epsilon+v-k)^2}{k(v-k)(e^\epsilon-1)^2 v}, \label{eq:OptimRiskAndSSRisk}
\end{equation}
respectively, where $k\in \{1,2,\cdots,v-1\}$ is the subset size parameter for the SS scheme and $k^* \in K^*_{v,\epsilon}$ is the optimal subset size. For $v \rightarrow \infty$, by applying  $k^* \approx \frac{v}{e^\epsilon+1}$, we can approximate $M_{v, 2,\epsilon}$ as follows: 
\begin{equation}
   M_{v, 2,\epsilon} \approx \frac{4ve^\epsilon}{(e^\epsilon-1)^2}. \label{eq:OptAsymRiskApprox}
\end{equation}

To the best of our knowledge, the SS scheme is the only known scheme which has been shown to be exactly optimal for every $(v,\epsilon)$. However, it requires a large communication cost in general, larger than the exponential growth in input data size. We note that RR\cite{RR-65Stanley} is a special case of SS corresponding to the subset size parameter $k=1$, and hence RR is exactly optimal for the privacy regime $\epsilon \geq \log(v-1)$, where $1 \in K^*_{v,\epsilon}$. Also, RAPPOR \cite{RAPPOR-14Erlingsson} was shown to achieve near-optimal risk in the high privacy regime $\epsilon \approx 0$\cite{SS-18MinYe}. More precisely, RAPPOR satisfies
\begin{equation}
    {R}_{v,2} (Q, \hat{P}_n) = \frac{v-1}{v}\left(1+\frac{v^2 e^{\epsilon/2}}{(v-1)(e^{\epsilon/2}-1)^2}\right).
\end{equation}
As $v \rightarrow \infty$ and $\epsilon \rightarrow 0$, both ${R}_{v,2} (Q, \hat{P}_n)$ achieved by RAPPOR and $M_{v,2,\epsilon}$ can be approximated as $\frac{4v}{\epsilon^2}$. However, RAPPOR also suffers from the large communication cost, the exponential growth in the input data size.

To remedy the communication cost issue, there has been a growing interest in developing communication-efficient LDP schemes \cite{HR-19Acharya, PGR-22Feldman}.\footnote{{Recently, to reduce communication cost, using asymmetric schemes, i.e., each client can use different privacy mechanisms, or utilizing shared randomness between clients and server  has been actively studied\cite{RHR-20WeiNingChen, MMRC-22Shah, one_bit_LDP-15Bassily}. In this paper, however,  we focus on using symmetric schemes, i.e., each client uses the same privacy mechanism, in the absence of shared randomness. Thus, in the following discussion on related works, we only consider those works using symmetric schemes in the absence of shared randomness.}} Among them, a remarkable result is the Hadamard response \cite{HR-19Acharya}. It presents a baseline scheme and a generalized scheme, where the latter scheme corresponds to a generalized version of the former to improve the risk. For distinction, we refer HR to only the baseline scheme, and refer generalized HR (GHR) to the generalized scheme. The GHR corresponds to an interpolation between the HR and the RR, and the optimized GHR is reduced to HR and a slight variant of RR in high and low privacy regimes, respectively. The GHR uses a small communication cost of $v+1 \leq b \leq 2v$, i.e., at most 1-bit more than the input data size for each client. However, it does not exactly achieve $M_{v,2,\epsilon}$ in general. More precisely, the HR achieves  (see \cite[equation (11)]{HR-19Acharya})
\begin{equation}
    {R}_{v,2} (Q, \hat{P}_n) \leq \frac{4v(e^\epsilon+1)^2}{(e^\epsilon - 1)^2}, \label{eq:HRRiskUppBd}
\end{equation}
and the GHR achieves (see \cite[equation (16) of supplementary]{HR-19Acharya}, or \cite[Table 1]{PGR-22Feldman})
\begin{equation}
    {R}_{v,2} (Q, \hat{P}_n) \leq \frac{36e^\epsilon(v+(e^\epsilon-1)b')}{(e^\epsilon - 1)^2}, \label{eq:GHRRiskUppBd}
\end{equation}
{where $b'=2^{\left\lceil \log_2 \left(\frac{v}{B}+1 \right) \right\rceil}$, $B=2^{\left\lceil\log_2 \min\{e^\epsilon, 2v\} \right\rceil - 1}$.
As $v \rightarrow \infty$ and $\epsilon \rightarrow 0$, the upper bound in \eqref{eq:HRRiskUppBd} can be approximated as $\frac{4v(e^\epsilon+1)^2}{(e^\epsilon - 1)^2} \approx \frac{16v}{\epsilon^2}$, and {as $v \rightarrow \infty$} the upper bound in \eqref{eq:GHRRiskUppBd} can be approximated as $\frac{36e^\epsilon(v+(e^\epsilon-1)b')}{(e^\epsilon - 1)^2} \approx \frac{36ve^\epsilon}{(e^\epsilon-1)^2}$, which are about 4-9 times larger than the optimal asymptotic risk \eqref{eq:OptAsymRiskApprox}. 
The risk over $\ell_1$ loss can be similarly analyzed.}
In addition, \cite{HR-19Acharya} conducted empirical evaluations for the GHR.
It was shown that the empirical risk of the GHR is close to that of optimal SS in high privacy regime.
However, in medium privacy level, the GHR shows about 10\% larger $\ell_2^2$ and $\ell_1^1$ empirical risks than the optimal SS.
Another experiment by \cite{PGR-22Feldman}  shows that GHR in medium privacy regime has about $2$ times larger empirical risk than that of optimal SS.
Note that in low privacy regime GHR achieves {near-optimal} risk as it is reduced to a variant of RR.
{Another recently proposed communication-efficient LDP scheme is the projective geometry response (PGR) \cite{PGR-22Feldman}. This is based on an observation that a similar structural properties used in construction of HR can be extracted also from a special mathematical object called a projective space. In fact, it was shown in \cite[Lemma 3.2]{PGR-22Feldman} that PGR achieves ${R}_{v,2} (Q, \hat{P}_n) \leq \frac{4ve^\epsilon}{(e^\epsilon-1)^2}$, which coincide with the approximated value of $M_{v,2,\epsilon}$ in \eqref{eq:OptAsymRiskApprox}, for a certain class of privacy regimes, i.e., $e^\epsilon+1 = q$ for some prime power $q$, while consuming little communication cost of $v \leq b \leq qv+1$. }

{In summary, some communication-efficient LDP schemes were shown to have either analytical or experimental evidence of \emph{approximately} achieving the optimal asymptotic risk in certain privacy regimes, where the approximation becomes more accurate as the corresponding limits approach, i.e., $v\rightarrow \infty$, and/or $\epsilon\rightarrow 0$ or $\epsilon\rightarrow \infty$. However, it is still not known whether these schemes are \emph{exactly} optimal. In this paper, we propose a new class of communication-efficient LDP schemes based on combinatorial block designs \cite{CombSymmDesigns-06Ionin} for discrete distribution estimation, called block design schemes. Our work distinguishes itself from previous studies on communication-efficient LDP schemes by focusing on the exact optimality without requirements for taking a limit on $v$ or $\epsilon$.}

\section{Main Results}\label{sec:mainSection}

\subsection{Preliminary: Incidence structure and block design}\label{subsec:blockDesign}
In this subsection, we introduce the basic definitions about the block design and its properties. The details can be found in \cite{CombSymmDesigns-06Ionin, ExtremalCombinatorics-11Jukna}.
\begin{definition}
An \textbf{incidence structure} is a triplet $(\mathcal{X},\mathcal{Y},\mathcal{I})$, where
\begin{itemize}
    \item $\mathcal{X}$, $\mathcal{Y}$ are sets.
    \item $\mathcal{I} \subset \mathcal{X} \times \mathcal{Y}$ is a subset of pairs $(x,y)$, $x \in \mathcal{X}$, $y \in \mathcal{Y}$.
\end{itemize}
If there is no confusion on $\mathcal{X}$ and $\mathcal{Y}$, we simply say $\mathcal{I}$ is an incidence structure. Furthermore, we define 
\begin{IEEEeqnarray}{rCl}
    \mathcal{I}_x &:=& \{y \in \mathcal{Y} : (x,y) \in \mathcal{I}\},\\
    \mathcal{I}^y &:=& \{x \in \mathcal{X} : (x,y) \in \mathcal{I}\}. 
\end{IEEEeqnarray}
\end{definition}

\begin{definition}
Let $(\mathcal{X},\mathcal{Y},\mathcal{I})$ be an incidence structure with $|\mathcal{X}|,|\mathcal{Y}|<\infty$. We say this incidence structure is
\begin{itemize}
    \item $r$-regular, if $|\mathcal{I}_x|=r$ for all $x \in \mathcal{X}$.
    \item $k$-uniform, if $|\mathcal{I}^y|=k$ for all $y \in \mathcal{Y}$.
    \item $\lambda$-{pairwise balanced}, if $|\mathcal{I}_{x} \cap \mathcal{I}_{x'}|=\lambda$ for all $x,x' \in \mathcal{X}$, $x \neq x'$.
\end{itemize}    
\end{definition}
\begin{definition}
Let $v,b,r,k$, and $\lambda$ be integers such that ${v>k>0}$, $b>r>\lambda \geq 0$. A \textbf{block design} with parameters $(v,b,r,k,\lambda)$, or shortly $(v,b,r,k,\lambda)$-block design, is an incidence structure such that
\begin{itemize}
    \item $\mathcal{X},\mathcal{Y}$ are finite sets with $|\mathcal{X}|=v$, $|\mathcal{Y}|=b$.
    \item $(\mathcal{X},\mathcal{Y},\mathcal{I})$ is $r$-regular, $k$-uniform, and $\lambda$-pairwise balanced.
\end{itemize}
\end{definition}
\begin{lemma}\cite[Proposition 2.3.7]{CombSymmDesigns-06Ionin}\label{lem:blockDesignFundEq}
For any block design, the parameters $(v,b,r,k,\lambda)$ should satisfy the following two equalities:
\begin{itemize}
    \item $vr=bk$.
    \item $\lambda(v-1)=r(k-1)$.
\end{itemize}
\end{lemma}

\subsection{Block design schemes: Locally differentially private schemes based on block designs}\label{subsec:BDM}
We propose a general class of schemes based on block designs, which not only contains randomized response (RR) \cite{RR-65Stanley}, subset selection (SS) \cite{SS-18MinYe}, Hadamard response (HR) \cite{HR-19Acharya}, and projective geometry response (PGR) \cite{PGR-22Feldman}, but also contains novel schemes achieving {exact optimality} while requiring lower communication costs. Our scheme is motivated by the observation that the aforementioned previously known mechanisms can be stated in the following way: 
 \begin{enumerate}
     \item An incidence structure $(\mathcal{X},\mathcal{Y},\mathcal{I})$, which is $r$-regular, $k$-uniform, and $\lambda$-pairwise balanced for some $r$, $k$, and $\lambda$, is determined in a certain way, which is different for each mechanism.
     \item For given $X=x$, the mechanism draws an independent Bernoulli random variable $B$.
     The mechanism outputs an element in $\mathcal{I}_x$ uniform randomly if $B=1$, and else, outputs an element in $\mathcal{Y} \setminus \mathcal{I}_x$ uniform randomly. 
 \end{enumerate}

  \begin{remark}\label{rmk:HRFramework}
 The works \cite{HR-19Acharya}, \cite{PGR-22Feldman} mentioned that their mechanisms satisfy the regular and the pairwise balanced properties{, and presented a framework of privacy schemes using these two properties}. We note that, however, they described these properties not in the context of incidence structure or block design, and did not capture the uniformity.
 \end{remark}

 \begin{remark}\label{rmk:HRModification}
     Hadamard response in \cite{HR-19Acharya} considers some truncated submatrix of Hadamard matrix when there is no Hadamard matrix matching to the input data size. For simplicity, we only refer to HR without such truncation, and we introduce the concept of truncation in Section \ref{subsec:RPBD}. Also, the Hadamard matrix used for the original version of HR does not satisfy uniformity. But, its slight modification by removing the first column and the first row satisfies uniformity and the corresponding scheme achieves lower worst-case $\ell_2^2$ risk than the original version.
    In this paper, we refer to the HR with this modification.
 \end{remark}

Now, let us introduce our proposed class of schemes. We first present the proposed mechanisms. 
{
\begin{definition}\label{def:blockDesignMech}
Let $\mathcal{X}=\{1,2,\cdots,v\}$ and $\epsilon>0$ be given. A \textbf{block design mechanism} is a privacy mechanism $Q$ which can be constructed as follows:
\begin{itemize}
    \item First, choose a block design $(\mathcal{X},\mathcal{Y},\mathcal{I})$, and
    \item Set $Q:\mathcal{X} \rightarrow \mathcal{Y}$ as
    \begin{equation}\label{eq:mech}
        Q(y|x)=\begin{cases} \alpha e^\epsilon & ((x,y) \in \mathcal{I}) \\ \alpha & ((x,y) \notin \mathcal{I}) \end{cases},
    \end{equation}
    where $\alpha = \frac{1}{r e^\epsilon + b - r}$ is a normalization constant.
\end{itemize}    
{If $Q$ is constructed from a block design $(\mathcal{X},\mathcal{Y},\mathcal{I})$ with parameters $(v,b,r,k,\lambda)$, then we say that such $Q$ is a $(v,b,r,k,\lambda,\epsilon)$-block design mechanism.}
\end{definition}
}

For matching estimators,  we construct a canonical unbiased estimator for each block design mechanism in a similar way as in \cite{SS-18MinYe,HR-19Acharya,PGR-22Feldman}, which is a simple linear function of the counts of the events $Y_i \in \mathcal{I}_x$. 
For simplicity, we call a pair of a block design mechanism and the corresponding canonical estimator as a \textbf{block design scheme}. 
\begin{theorem}\label{thm:BDMEstimator}
Given $n$ and a $(v,b,r,k,\lambda,\epsilon)$-block design mechanism $Q$, there is an unbiased\footnote{Unbiasedness means $\mathbb{E}_{P,Q}[\hat{P}_n]=P$ for all $P \in \Delta_v$} estimator given as
\begin{align}\label{eq:Canonical est}
    \hat{P}_{n,x}(Y_1,\cdots,Y_n) = \frac{1}{(r-\lambda)(e^\epsilon-1)}
     \left( \frac{N_x(Y_1,\cdots,Y_n)}{n\alpha} -(\lambda e^\epsilon+(r-\lambda)) \right),
\end{align}
where
\begin{align}\label{eq:Nx}
N_x(Y_1,\cdots,Y_n) = \sum_{i=1}^{n}\mathbbm{1}(Y_i \in \mathcal{I}_x).
\end{align}
\end{theorem}

There are many known block designs, all of which can be used to construct LDP schemes.
For example, \cite{Colbourn96-CRCCombDesignsHandbook} presents more than a thousand of block designs and lots of structured families of block designs throughout the book.
In Table \ref{tab:blockDesigns}, we present some represented block designs based on special mathematical structures\cite{CombSymmDesigns-06Ionin, DifferenceSets-13Moore}.
We note that many previously proposed schemes are special cases of block design schemes as indicated in Table \ref{tab:blockDesigns}. 
Furthermore, for some block designs induced from special mathematical structures, we can also present efficient algorithms for the corresponding block design schemes.
As an illustration, in Appendix \ref{app:blockDesignsDescription} of the supplementary material, we present a detailed description of block designs in Table \ref{tab:blockDesigns}, and in Appendix \ref{app:blockDesignsAlgorithms} of the supplementary material, we present some efficient algorithms for implementing block design schemes derived from block designs in Table \ref{tab:blockDesigns} that are not previously studied. 

\begin{table*}[ht]
    \centering
    \caption{List of some block designs\cite{CombSymmDesigns-06Ionin, DifferenceSets-13Moore}}
    \label{tab:blockDesigns}
    \begin{tabular}{c|c|c|c|c|c|c|c}
        & & & & & & & Corresponding\\ 
        Structure & Parameters & $v$ & $b$ & $r$ & $k$ & $\lambda$ & previous \\
       & & & & & & & schemes\\
        \hline\hline
        \makecell{Trivial symmetric\\block design} & $v$ : An integer $v \geq 2$ & $v$ & $v$ & $1$ & $1$ & $0$ & RR\cite{RR-65Stanley}\\
        \hline
        \makecell{Complete\\block design} & \makecell{$v$ : An integer $v \geq 2$\\$k$ : An integer $1 \leq k \leq v-1$} & $v$ & $\binom{v}{k}$ & $\binom{v-1}{k-1}$ & $k$ & $\binom{v-2}{k-2}$ & SS\cite{SS-18MinYe}\\
        \hline
        \makecell{Projective geometry} & \makecell{$q$ : prime power\\$t$ : An integer $t \geq 2$} &  $\frac{q^t-1}{q-1}$ & $\frac{q^t-1}{q-1}$ & $\frac{q^{t-1}-1}{q-1}$ & $\frac{q^{t-1}-1}{q-1}$ & $\frac{q^{t-2}-1}{q-1}$ & PGR\cite{PGR-22Feldman}\\
        \hline
        \makecell{Sylvester's\\Hadamard matrix} & \makecell{$t$ : An integer $t \geq 2$} & $2^t-1$ & \multirow{3}{*}[-1em]{$v$} & \multirow{3}{*}[-1em]{$(v-1)/2$} & \multirow{3}{*}[-1em]{$(v-1)/2$} & \multirow{3}{*}[-1em]{$(v-3)/4$} & HR\cite{HR-19Acharya}\\ \cline{1-3} \cline{8-8}
        
        \makecell{Paley's\\Hadamard matrix} & \makecell{$v$ : A prime power\\s. t. $v \equiv 3 \mod 4$} & $v$ & &  &  & & -\\ \cline{1-3} \cline{8-8}
        
        \makecell{Twin prime power\\difference set} & \makecell{$q$ : An odd prime power s.t. \\$q+2$ is also prime power} & $q(q+2)$ & & & & & -\\
        \hline
        \makecell{Nonzero\\Quartic residue\\difference set} & \makecell{$v$ : A prime power s.t. \\$v=4t^2+1$ for some\\odd integer $t$} & $v$ & $v$ & $(v-1)/4$ & $(v-1)/4$ & $(v-5)/16$ & -\\
        \hline
        \makecell{(Including zero)\\Quartic residue\\difference set} & \makecell{$v$ : A prime power s.t.\\$v=4t^2+9$ for some\\odd integer $t$} & $v$ & $v$ & $(v+3)/4$ & $(v+3)/4$ & $(v+3)/16$ & -
    \end{tabular}
\end{table*}

The risk and the worst case risk under mean squared loss of a block design scheme are given as follows. The proof is in Appendix \ref{app:proofs} of the supplementary material.
\begin{theorem}\label{thm:BDML2risk}
The risk under $\ell_2^2$  loss of a block design scheme is given by 
\begin{align}
    R_{v,2}^n(P, Q, \hat{P}_n)
    =\frac{1}{n}\left( \frac{(v-1)^2(ke^\epsilon+v-k)^2}{k(v-k)(e^\epsilon-1)^2 v} + \frac{1}{v}-\sum_{x \in \mathcal{X}}P_x^2\right),
\end{align}
and the worst-case risk is given by
\begin{IEEEeqnarray}{c}
    R_{v,2}^n(Q, \hat{P}_n)=\frac{(v-1)^2(ke^\epsilon+v-k)^2}{k(v-k)(e^\epsilon-1)^2 nv}, \label{eq:BDWorstCaseRisk}
\end{IEEEeqnarray}
which is attained when $P$ is a uniform distribution.
\end{theorem}
{
Especially, we have
\begin{equation}
    {R}_{v,2} (Q, \hat{P}_n) = \frac{(v-1)^2(ke^\epsilon+v-k)^2}{k(v-k)(e^\epsilon-1)^2 v}. \label{eq:BDAsympRisk}
\end{equation}
}
Note that for any given $n$, the risk under $\ell_2^2$ loss of a block design scheme only depends on $(v,k,\epsilon)$. In other words, for given input data size $v$ and LDP constraint $\epsilon$, the risk of a block design scheme with parameter $(v,b,r,k,\lambda,\epsilon)$ only depends on $k$ under $\ell_2^2$ loss. {Especially, the asymptotic risk in \eqref{eq:BDAsympRisk} coincide with that of SS scheme with the subset size parameter $k$ presented in \eqref{eq:OptimRiskAndSSRisk}.} We show that this is true for a broad class of loss functions including the $\ell_u^u$ loss, $u \geq 1$, by defining a kind of equivalence between two schemes as follows.  
\begin{definition}
Given $v$ and $n$, two schemes $(Q, \hat{P}_n)$ and $(Q',\hat{P}_n ')$ are said to be \textbf{marginally equivalent} if the marginal distributions of the two estimators given the distribution on $\mathcal{X}$ are the same.
Precisely, let $X_1,\cdots,X_n \sim P \in \Delta_v$, and $Y_i$ and $Y_i'$ be the outputs of the mechanisms $Q$ and $Q'$ for an input data $X_i$, respectively.
We say the two schemes are marginally equivalent if
\begin{IEEEeqnarray}{c}
    \hat{P}_{n,x}(Y_1,\cdots,Y_n) \overset{d}{=} \hat{P}_{n,x}'(Y_1',\cdots,Y_n'),
\end{IEEEeqnarray}
for all $x \in \mathcal{X}$ and $P \in \Delta_v$.
\end{definition}

Next, for given $(v,n,\epsilon)$, we show that the marginal equivalence of two block design schemes is determined solely by $k$ in the following theorem, whose proof is in Appendix \ref{app:proofs} of the supplementary material.
\begin{theorem}\label{thm:blockDesignEquiv}
    For given $(v,n,\epsilon)$, let $(Q,\hat{P}_n)$ and $(Q',\hat{P}_n')$ be block design schemes with parameters $(v, b, r,k, \lambda, \epsilon)$ and $(v, b',r',k',\lambda', \epsilon)$, respectively. Then, the two schemes are marginally equivalent if and only if $k=k'$.
\end{theorem}

Now, we present a class of widely-used loss functions which give the same risk for marginally equivalent schemes.
\begin{definition}
A loss function $\ell$ is said to be \textbf{decomposable} if there are functions $\ell_x:[0,1] \times \mathbb{R} \rightarrow \mathbb{R}$, $x=1,2,\cdots,v$, such that
\begin{align}
    \ell(p, \hat{p}) = \sum_{x=1}^{v} \ell_x(p_x, \hat{p}_x).
\end{align}
for all $p \in \Delta_v$, $\hat{p} \in \mathbb{R}^v$.
\end{definition}

For example, for each $u \geq 1$, $\ell_u^u$ loss is decomposable with $\ell_x(p_x,\hat{p}_x) = |p_x-\hat{p}_x|^u$. Another examples are $f$-divergences $D_f:\Delta_v \times \Delta_v \rightarrow \mathbb{R}$,
\begin{align}
D_f(\hat{p} \Vert p) = \sum_{x=1}^{v}p_x f(\hat{p}_x/p_x)
\end{align}
for convex functions $f:[0,\infty] \rightarrow \mathbb{R} \cup \{\infty\}$ such that ${f(1)=0}$.
It can be easily observed that marginally equivalent schemes give the same risk for a decomposable loss function.

\begin{proposition}\label{prop:equivSameRisk}
If $(Q, \hat{P}_n)$ and $(Q',\hat{P}_n ')$ are marginally equivalent, then 
\begin{equation}
    R_{v,\ell}^n(P, Q, \hat{P}_n)=R_{v,\ell}^n(P, Q', \hat{P}_n'),
\end{equation}
 for any $P \in \Delta_v$ and a decomposable loss function $\ell$.
\end{proposition}
\begin{IEEEproof}
We have
\begin{align}
    R_{v,\ell}^n(P, Q, \hat{P}_n)
     = \sum_{x=1}^{v} \expect\left[ \ell_x\left( P_x,\hat{P}_{n,x}(Y_1,\cdots,Y_n) \right)\right].
\end{align}
Each of $\expect\left[ \ell_x\left( P_x,\hat{P}_{n,x}(Y_1,\cdots,Y_n) \right)\right]$ only depends on the marginal distribution of each $\hat{P}_{n,x}(Y_1,\cdots,Y_n)$, not on the joint distribution of $\hat{P}_n(Y_1,\cdots,Y_n)$. Thus, marginally equivalent schemes give the same risk. 
\end{IEEEproof}

Now, together with the {exact optimality} of SS \cite{SSOptim-19MinYe}, we prove the {exact optimality} of a block design scheme.
{
\begin{theorem}\label{thm:optimBlockDesign}
    For given $(v,\epsilon)$, let $(Q,\hat{P}_n)$ be a block design scheme satisfying $k \in K^*_{v,\epsilon}$, where
    \begin{equation}
        K^*_{v,\epsilon} = \argmin_{k \in \mathbb{Z}, 1 \leq k \leq v-1} \frac{(ke^\epsilon+v-k)^2}{k(v-k)}
    \end{equation}
    as defined in Section \ref{subsec:prevWorks}. Then $(Q,\hat{P}_n)$ is exactly optimal for any $1 \leq u \leq 2$, that is
    \begin{IEEEeqnarray}{c}\label{eq:SSoptim}
        {R}_{v,u} (Q, \hat{P}_n)=M_{v, u,\epsilon}
    \end{IEEEeqnarray}
    for all $1 \leq u \leq 2$.
\end{theorem}
}
\begin{IEEEproof}
{As presented in Theorem \ref{thm:SSOptim},} \cite[Theorems 3.2 and 3.3]{SSOptim-19MinYe} showed that (\ref{eq:SSoptim}) holds when $(Q,\hat{P}_n)$ is the SS scheme with $k \in K^*_{v,\epsilon}$.
Also, Theorem \ref{thm:blockDesignEquiv} and Proposition \ref{prop:equivSameRisk} say that any block design schemes with the same $k$ induce the same risk.
Hence, \eqref{eq:SSoptim} also holds when we replace $(Q,\hat{P}_n)$ by any block design scheme with the same $k$.
\end{IEEEproof}

We also strengthen the characterization of $K^*_{v,\epsilon}$ presented in \eqref{eq:optimUnifParam} into the following proposition.
\begin{proposition}\label{prop:optEpsRegion}
For given $(v,k,\epsilon)$, $k \in K^*_{v,\epsilon}$ holds if and only if
\begin{align}
    E(k, k+1;v) \leq e^\epsilon \leq E(k-1,k;v),
\end{align}
where
\begin{align}
    E(k_1,k_2;v)=\sqrt{\frac{(v-k_1)(v-k_2)}{k_1 k_2}},
\end{align}
and $E(0,1;v):=\infty$.
Especially, if $e^\epsilon=E(k,k+1;v)$ for some $k$, then $K^*_{v,\epsilon}=\{k,k+1\}$. Otherwise, $K^*_{v,\epsilon}$ is a singletone.
\end{proposition}
\begin{IEEEproof}
Let us recall that
\begin{align}
    E(k_1,k_2;v)=\sqrt{\frac{(v-k_1)(v-k_2)}{k_1 k_2}},
\end{align}
which is defined for $0 \leq k_1 < k_2 \leq v$, $(k_1,k_2) \neq (0,v)$. We set $E(0, k_2;v) = \infty$ for any $k_2 \in [1:v-1]$.

Let $L(v,k,\epsilon) = \frac{(ke^\epsilon+v-k)^2}{k(v-k)}$, $1 \leq k \leq v-1$. Also, let $L(v,0,\epsilon)=L(v,v,\epsilon)=\infty$. First, let us present the following lemma, whose proof will be presented right after the proof of this proposition.
\begin{lemma} \label{lemmaA}
Let $0 \leq k_1 < k_2 \leq v$, $(k_1,k_2) \neq (0,v)$. Then $L(v,k_1,\epsilon) \leq L(v,k_2,\epsilon)$ if and only if $e^\epsilon \geq E(k_1,k_2;v)$.
\end{lemma}
First assume that $k \in K^*_{v,\epsilon}$. Then, $L(v,k,\epsilon) \leq L(v,k-1,\epsilon)$ and $L(v,k,\epsilon) \leq L(v,k+1,\epsilon)$.
Then, by the above lemma, we have $E(k, k+1;v) \leq e^\epsilon \leq E(k-1,k;v)$.

Conversely, assume that $E(k, k+1;v) \leq e^\epsilon \leq {E(k-1,k;v)}$. Note that $E(k_1,k_2;v)$ is strictly decreasing in both $k_1$ and $k_2$. Thus, we have $E(k,k';v) \leq E(k,k+1;v) \leq e^\epsilon$ for all $k' \geq k+1$. Thus, by the lemma, $L(v,k,\epsilon) \leq L(v,k',\epsilon)$ for all $k' \geq k+1$. Similarly, $E(k',k;v) \geq E(k-1,k;v) \geq e^\epsilon$ for all $k' \leq k-1$. Thus, $L(v,k,\epsilon) \leq L(v,k',\epsilon)$ for all $k' \leq k-1$. This shows that given $k$ minimizes $L(v,k,\epsilon)$, so that $k \in K^*_{v,\epsilon}$. 

The statement about the cardinality of $K^*_{v,\epsilon}$ directly follows from the observation that $E(k, k+1;v)$ is strictly decreasing in $k$.
\end{IEEEproof}

\begin{IEEEproof}[Proof of Lemma \ref{lemmaA}]
If $k_1=0$, then ${L(v,k_1,\epsilon)=\infty}$, ${L(v,k_2,\epsilon)<\infty}$, $E(k_1,k_2;v)=\infty$. Hence, the statement is vacuously true. Also, if $k_2=v$, then $L(v,k_2,\epsilon)=\infty$, $E(k_1,k_2;v)=0$. Hence, the statement is always true. Thus, it suffices to show for $k_1,k_2 \in [1:v-1]$.

We can observe that
\begin{align}
    \sqrt{L(v,k,\epsilon)} &= \frac{ke^\epsilon+v-k}{\sqrt{k(v-k)}}
    \\ &= \sqrt{\frac{k}{v-k}}e^\epsilon + \sqrt{\frac{v-k}{k}}.
\end{align}
Let $C_1 = \sqrt{\frac{v-k_1}{k_1}}$, $C_2 = \sqrt{\frac{v-k_2}{k_2}}$. Note that $C_1 > C_2$ and $C_1 C_2 = E(k_1,k_2 ; v)$. Then, the condition $L(v,k_1,\epsilon) \leq L(v,k_2,\epsilon)$ can be equivalently written as
\begin{gather}
    \sqrt{L(v,k_1,\epsilon)} \leq \sqrt{L(v,k_2,\epsilon)}\\
    \Leftrightarrow \frac{1}{C_1}e^\epsilon +C_1 \leq \frac{1}{C_2}e^\epsilon +C_2\\
    \Leftrightarrow \frac{C_2 - C_1}{C_1 C_2} e^\epsilon \leq C_2 - C_1\\
    \Leftrightarrow e^\epsilon \geq C_1 C_2 = E(k_1,k_2 ; v).
\end{gather}
This ends the proof of the lemma.
\end{IEEEproof}

{
It should be noted that the exact optimality as in Theorem \ref{thm:optimBlockDesign} holds among all schemes, not necessarily unbiased ones. Also, the exact optimality holds for any $\ell_u^u$ loss, $1 \leq u \leq 2$. However, if we restrict our attention to only unbiased schemes and $\ell_2^2$ loss, then we can get a more stronger result. We show that our scheme has the smallest worst-case risk among all unbiased LDP schemes for each fixed $n$, without involving any asymptotic statement. {We note that such a non-asymptotic optimality was also considered in \cite{PrivUnitOptim-22Asi} but for a different task of mean estimation.}
}
\begin{theorem}\label{thm:nonasymp_opt}
    For given $n$, let $(Q,\hat{P}_n)$ be a block design scheme satisfying $k \in K^*_{v,\epsilon}$, and $(Q', \hat{P}_n')$ be any unbiased $\epsilon$-LDP scheme.
    Then, we have
    \begin{IEEEeqnarray}{c}
        R_{v,2}^n(Q, \hat{P}_n) \leq R_{v,2}^n(Q', \hat{P}_n').
    \end{IEEEeqnarray}
\end{theorem}
{
\begin{IEEEproof}
For each positive integer $m \geq n$, let $\hat{P}'_{m}$ be the estimator on $m$ clients given by
    \begin{align}
    \hat{P}'_{m}(Y_1',Y_2',\cdots,Y_{m}')
    = \frac{1}{\lfloor m/n \rfloor}\sum_{j=0}^{\lfloor m/n \rfloor-1} \hat{P}_n' (Y_{nj+1}',Y_{nj+2}',\cdots,Y_{nj+n}').
    \end{align}
    Because $Y_1',Y_2',\cdots,Y_{m}'$ are i.i.d. and $(Q',\hat{P}_n')$ is unbiased, $(Q',\hat{P}_{m}')$ is also unbiased.
    Thus, the risk under the mean squared loss is same as the variance of the estimator.
    Therefore,
    \begin{IEEEeqnarray}{c}
        R_{v,2}^m(P, Q', \hat{P}'_{m}) = \frac{1}{\lfloor m/n \rfloor}R_{v,2}^n(P, Q', \hat{P}_n'),
    \end{IEEEeqnarray}
    and consequently,
    \begin{IEEEeqnarray}{c}
        R_{v,2}^m(Q', \hat{P}'_{m}) = \frac{1}{\lfloor m/n \rfloor}R_{v,2}^n(Q', \hat{P}_n').
    \end{IEEEeqnarray}
 Also, from \eqref{eq:BDWorstCaseRisk}, it can be easily observed that
    \begin{equation}
        R_{v,2}^m(Q, \hat{P}_{m}) = \frac{1}{m/n}R_{v,2}^n(Q, \hat{P}_n).
    \end{equation}
    From this and Theorem \ref{thm:optimBlockDesign}, we have
    \begin{align}
    M_{v,2,\epsilon} &= {R}_{v,2} (Q, \hat{P}_m) = \lim_{m \rightarrow \infty} m {R}_{v,2}^m(Q, \hat{P}_{m}) = n {R}_{v,2}^n(Q, \hat{P}_n)\\
    & \leq {R}_{v,2} (Q, \hat{P}'_m) = \lim_{m \rightarrow \infty} m {R}_{v,2}^m(Q, \hat{P}'_{m}) = n {R}_{v,2}^n(Q, \hat{P}'_n)
    \end{align}
    where we use the fact that $\lim_{m \rightarrow \infty} \frac{m}{\lfloor m/n \rfloor} = n$. By cancelling out $n$ in both lines, we get the desired result.
\end{IEEEproof}
}

The consequences of our exact optimality results are as follows.
\begin{itemize}
    \item Although HR\cite{HR-19Acharya} and PGR\cite{PGR-22Feldman} have been shown to be near-optimal through either analytical or experimental results, it was not shown whether they are exactly optimal. Our result shows that in fact these two schemes are exactly optimal for specific regimes of $(v,\epsilon)$.
    \item By utilizing other block designs not previously used to construct an LDP scheme, we can construct exactly optimal schemes with low communication cost for many other regimes of $(v,\epsilon)$.
\end{itemize}


\subsection{Block design schemes with low communication costs} \label{subsec:symmetric}
Theorem \ref{thm:optimBlockDesign} shows that for any given $(v,\epsilon)$, a block design scheme satisfying $k \in K^*_{v,\epsilon}$ {is exactly optimal.}
There might be several different block design schemes with $k \in K^*_{v,\epsilon}$.
Among them, a block design scheme with $b=v$ uses the smallest communication cost: Fisher's inequality\cite{CombSymmDesigns-06Ionin} tells that any block design should satisfy $b \geq v$. Such block designs with $b=v$ are called \textbf{symmetric block designs}. Furthermore, we can prove that $b \geq v$ is also required for the {unbiasedness or consistency} of a scheme, in the same way to the proof of \cite[Thm. 6]{comm_unbiased-19Acharya}.
{
\begin{theorem}
Let $\mathcal{X}=\{1,2,\cdots,v\},\mathcal{Y}$ be a finite set, and ${Q:\mathcal{X} \rightarrow \mathcal{Y}}$ be any conditional distribution.
If $(Q,\hat{P}_n)$ is an unbiased or consistent scheme\footnote{Consistency means that for any fixed $P$, we have $\hat{P}_n \rightarrow P$ in probability. Note that inconsistent scheme always has non-vanishing worst-case risk as $n \rightarrow \infty$ over any $\ell_u^u$ loss, since $\ell_u$ convergence implies convergence in probability.}, then $|\mathcal{Y}| \geq v$.
\end{theorem}
}

\begin{figure}[htbp]
    \begin{center}
        \includegraphics[width=0.5\textwidth]{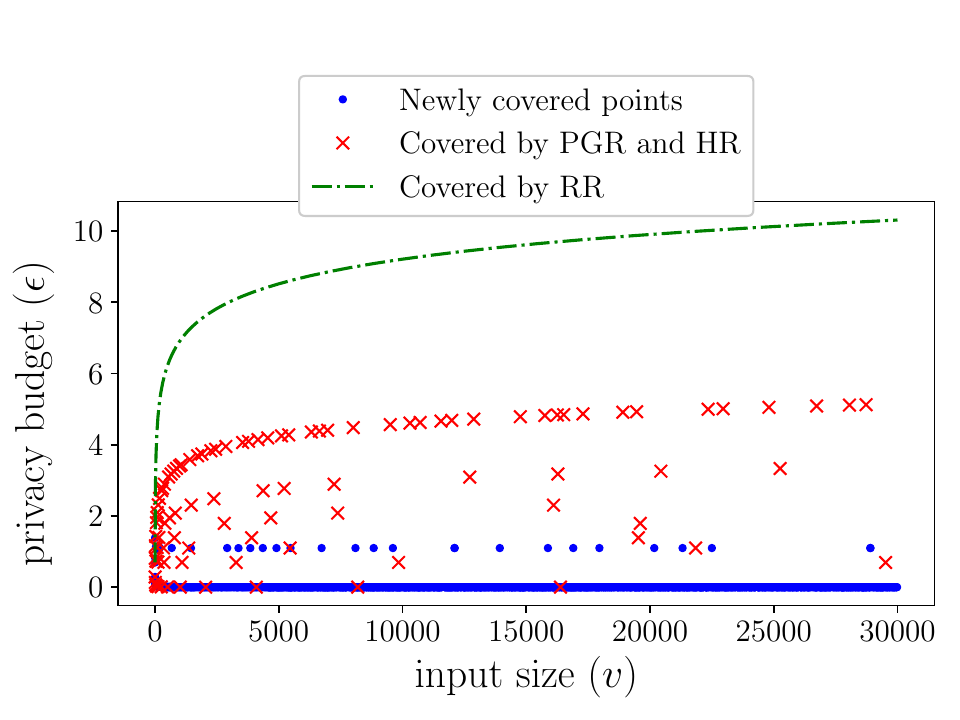}
    \end{center}
    \caption{The points of $(v,\epsilon)$ for which both the {exactly optimality} and the minimum communication cost are achievable by using the symmetric block designs listed in Table \ref{tab:blockDesigns}. }
    \label{fig:optRegion}
\end{figure}

Thus, a symmetric block design scheme with $k \in K^*_{v,\epsilon}$ is an unbiased and consistent scheme which achieves the exact optimality while using the minimum communication cost for unbiasedness or consistency.
We note that the block designs in Table \ref{tab:blockDesigns} except the complete block design are all symmetric block designs. 
{In Fig. \ref{fig:optRegion}, we plot the points of $(v,\epsilon)$ for which both the {exact optimality} and the minimum communication cost are achievable by using the symmetric block designs listed in Table \ref{tab:blockDesigns}. 
For simplicity, we only draw points $(v,\epsilon=\log((v/k)-1))$ for $v$ and $k$ specified in Table \ref{tab:blockDesigns}, for which $k \in K^*_{v,\epsilon}$ trivially holds.\footnote{In fact, the {exact optimality} and the minimum communication cost are obtained when $\epsilon$ is in a neighborhood $[\log E(k, k+1;v), \log E(k-1,k;v)]$ of $\log((v/k)-1))$ as stated in Proposition \ref{prop:optEpsRegion}.}
In the figure, the red crosses are the points where the {exact optimality is} achievable by HR \cite{HR-19Acharya} or PGR \cite{PGR-22Feldman}, green dash-dot line by RR \cite{RR-65Stanley}, and blue dots by newly covered points by using other symmetric block designs in Table \ref{tab:blockDesigns}. We group PGR and HR since HR is equivalent to PGR with $q=2$.
As shown in the figure, many new  points of $(v,\epsilon)$ are covered by considering only the four symmetric block designs not used in the previously proposed mechanisms in Table \ref{tab:blockDesigns}.
In particular, we can check that the points in high privacy ($\epsilon \approx 0$) are covered very densely.
The coverage in high privacy regime is related to the famous conjecture called the Hadamard conjecture\cite{HadamardConjSurvey-04Tressler}, which is discussed in Section \ref{subsec:hadConj}. Shortly saying, in high privacy regime, we cover very dense subset of $v$ among points of the form $v \equiv 3 \mod 4$, and the Hadamard conjecture implies that we can cover \emph{all} points of the form $v \equiv 3 \mod 4$.
We note that there are many other symmetric block designs not listed in Table \ref{tab:blockDesigns}, e.g., 23 families of symmetric block designs are listed in  \cite[Appendix]{CombSymmDesigns-06Ionin}. 
Accordingly, we can cover more points of $(v,\epsilon)$ by employing all of them.}

In general, however, a symmetric block design scheme may or may not exist for given $(v,k)$.
For example, if $v= 8$ and {$\epsilon = 1$}, then $K^*_{v,\epsilon}=\{2\}$. 
If we assume the existence of a symmetric block design with $v=b=8$ and $k=2$, then Lemma \ref{lem:blockDesignFundEq} says that $\lambda = 2/7 \notin \mathbb{Z}$, which gives a contradiction. 
Also, there are some other necessary conditions for the existence of a symmetric block design, for example the Bruck-Ryser-Chowla theorem\cite[Theorem 2.5.10]{CombSymmDesigns-06Ionin}.
If there is no symmetric block design with $k \in K^*_{v,\epsilon}$, {then a block design with $k \in K^*_{v,\epsilon}$, and reasonably low $b$ could be used to achieve the exact optimality with reasonably low communication cost.
Furthermore, if we also sacrifice the utility, i.e., targeting near-optimality instead of exact optimality, a block design with $k \approx v/(e^\epsilon+1)$ and reasonably low $b$ can be used.
In the following, we show an example of obtaining a non-symmetric block design achieving {exact or near-optimality} with reasonably low $b$.}
\begin{example}\label{ex:dereived}
One way of obtaining a non-symmetric block design with reasonably low $b$ is to utilize derived and residual block designs \cite{CombSymmDesigns-06Ionin}.
Let $(\mathcal{X},\mathcal{Y},\mathcal{I})$ be a $(v,b,r,k,\lambda)$-symmetric block design with $\lambda>0$, and let $y \in \mathcal{Y}$. Note that from Lemma \ref{lem:blockDesignFundEq} with $b=v$, we must have $r=k$ and $\lambda=k(k-1)/(v-1)$. 
Then the following two incidence structures $(\mathcal{X}', \mathcal{Y}', \mathcal{I}')$, $(\mathcal{X}'', \mathcal{Y}'', \mathcal{I}'')$ given by
\begin{gather}
    \mathcal{X}' = \mathcal{I}^y , \quad \mathcal{X}'' = \mathcal{X} \backslash \mathcal{I}^y,\\
    \mathcal{Y}' = \mathcal{Y}'' = \mathcal{Y} \backslash \{y\},\\
    \mathcal{I}' = \mathcal{I} \cap (\mathcal{X}' \times \mathcal{Y}'), \quad \mathcal{I}'' = \mathcal{I} \cap (\mathcal{X}'' \times \mathcal{Y}''),
\end{gather}
are $(k, v-1, k-1, \lambda, \lambda-1)$- and $(v-k,v-1,k,k-\lambda,\lambda)$-block designs, respectively, called the \textbf{derived block design} and the \textbf{residual block design}, respectively. The proof that they satisfy the properties of block design can be found in \cite[Proposition 2.4.15]{CombSymmDesigns-06Ionin}. Note that if $v/k \approx e^\epsilon+1$, then from Lemma \ref{lem:blockDesignFundEq}, the parameters $(v',b',r',k',\lambda')$ and $(v'',b'',r'',k'',\lambda'')$ of the derived and residual block designs satisfy
\begin{align}
    \frac{v'}{k'} = \frac{k}{\lambda} = \frac{v-1}{k-1} \approx \frac{v}{k} \approx e^\epsilon+1
\end{align}
and 
\begin{align}
    v''/k'' = \frac{v-k}{k-\lambda} = \frac{v-k}{k\left(1-\frac{k-1}{v-1}\right)} = \frac{v-1}{k} \approx \frac{v}{k} \approx e^\epsilon+1,
\end{align}
respectively, for $v,k \gg 1$.
{Hence, if a $(v,b,r,k,\lambda,\epsilon)$-symmetric block design scheme is {exactly or nearly optimal}, then the derived and residual block design schemes from the symmetric block design is exactly or nearly optimal for $v'=k$ and $v''=v-k$, respectively, for the same $\epsilon$.}
\end{example}

Up to now, we have considered block-design schemes which are exactly or nearly optimal with low communication costs. In the next subsection, to overcome the sparse existence of block designs due to some necessary conditions on the parameters, we consider a broader class of schemes by relaxing a constraint on the block design.

\subsection{RPBD schemes: A broader class of schemes based on regular and pairwise-balanced designs}\label{subsec:RPBD}

By carefully analyzing the block design schemes, it can be checked that the regular and pairwise balanced properties play crucial roles in their construction, while the uniformity does not.
Also, the framework in \cite{HR-19Acharya,PGR-22Feldman} used for HR and PGR does not consider uniformity in the construction of schemes.
Motivated from this, we propose a larger class of schemes than block design schemes by relaxing the uniformity constraint on the block designs. 
This class is also a generalization of the aforementioned framework of \cite{HR-19Acharya,PGR-22Feldman}.

\begin{definition}\label{def:RPB}
    Let $v,b,r$, and $\lambda$ be integers such that ${v>0}$, $b>r>\lambda \geq 0$.
    A \textbf{regular and pairwise-balanced design (RPBD)} with parameters $(v,b,r,\lambda)$, or in short a $(v,b,r,\lambda)$-RPBD, is an incidence structure $(\mathcal{X},\mathcal{Y},\mathcal{I})$, where
\begin{itemize}
    \item $\mathcal{X},\mathcal{Y}$ are non-empty finite sets with $|\mathcal{X}|=v$, $|\mathcal{Y}|=b$.
    \item $(\mathcal{X},\mathcal{Y},\mathcal{I})$ is $r$-regular and $\lambda$-pairwise balanced.
\end{itemize}
\end{definition}
\begin{remark} \label{remark:truncation}
    A typical example of an RPBD is obtained by truncating a block design. Given a $(v',b,r,k,\lambda)$-block design $(\mathcal{X}',\mathcal{Y},\mathcal{I}')$ and a subset $\mathcal{X} \subset \mathcal{X}'$ where $|\mathcal{X}|=v$, an incidence structure $(\mathcal{X},\mathcal{Y},\mathcal{I})$ given by
\begin{align}
    \mathcal{I} = \mathcal{I}' \cap (\mathcal{X} \times \mathcal{Y})
\end{align}
is a $(v,b,r,\lambda)$-RPBD, but not necessarily is a block design.
\end{remark}

Now, we define an RPBD scheme based on RPBD and present its risk.
\begin{definition}\label{def:RPB_mech}
For given $\mathcal{X}=\{1,2,\cdots,v\}$ and $\epsilon>0$, a privacy mechanism $Q$ is called an \textbf{RPBD mechanism} with parameter $(v,b,r,\lambda,\epsilon)$ if $Q$ is given as \eqref{eq:mech} for some $(v,b,r,\lambda)$-RPBD $(\mathcal{X},\mathcal{Y},\mathcal{I})$. 
\end{definition}

\begin{theorem}\label{thm:RPBL2risk}
Given $n$ and a $(v,b,r,\lambda,\epsilon)$-RPBD mechanism $Q$, there is an unbiased estimator given  as \eqref{eq:Canonical est}. A $(v,b,r,\lambda,\epsilon)$-RPBD mechanism and its corresponding unbiased estimator \eqref{eq:Canonical est} is called a $(v,b,r,\lambda,\epsilon)$-RPBD scheme. 
The risk of a $(v,b,r,\lambda,\epsilon)$-RPBD scheme under the mean squared loss is given by 

\begin{align} \label{eq:RPBL2risk}
        &R_{v, 2}^n(P, Q, \hat{P}_n) \\&= \frac{1}{n}\Biggl[\frac{\left[re^\epsilon+(v-1)(\lambda e^\epsilon+r-\lambda)\right]\left[v(b-r) + (v-1)(r-\lambda)(e^\epsilon-1) \right]}{(r-\lambda)^2(e^\epsilon-1)^2 v} +\frac{1}{v}-\sum_{x \in \mathcal{X}}P_x^2\Biggr],
\end{align}
and the worst-case risk is given by 
\begin{equation}\label{eq:RPBL2risk_opt}
        R_{v, 2}^n(Q, \hat{P}_n) = \frac{\left[re^\epsilon+(v-1)(\lambda e^\epsilon+r-\lambda)\right]\left[v(b-r) + (v-1)(r-\lambda)(e^\epsilon-1) \right]}{n(r-\lambda)^2(e^\epsilon-1)^2 v},
    \end{equation}
which is attained when $P$ is the uniform distribution.
\end{theorem}
The proof of the above theorem is in Appendix \ref{app:proofs} of the supplementary material. {Especially, a $(v,b,r,\lambda,\epsilon)$-RPBD scheme satisfies
\begin{equation}
    R_{v, 2}(Q, \hat{P}_n) = \frac{\left[re^\epsilon+(v-1)(\lambda e^\epsilon+r-\lambda)\right]\left[v(b-r) + (v-1)(r-\lambda)(e^\epsilon-1) \right]}{(r-\lambda)^2(e^\epsilon-1)^2 v}.
\end{equation}
}

Now, the following theorem gives a condition for the marginal equivalence of two RPBD schemes, whose proof is in  Appendix \ref{app:proofs} of the supplementary material.
\begin{theorem}\label{thm:RPBequiv}
    For given $(v,n,\epsilon)$, let $(Q,\hat{P}_n)$ and $(Q',\hat{P}_n')$ be two RPBD schemes with parameters $(v, b, r, \lambda, \epsilon)$ and $(v, b',r',\lambda', \epsilon)$, respectively. Then, the two schemes are marginally equivalent if and only if there exists a constant $t>0$ such that $(b',r',\lambda')=t(b,r,\lambda)$. In other words, the marginal equivalence is determined by the ratio $[b:r:\lambda]$.
\end{theorem}

We note that by Lemma \ref{lem:blockDesignFundEq} and the formula of $K^*_{v,\epsilon}$ in Proposition \ref{prop:optEpsRegion}, the optimal SS scheme has
\begin{align}
    [b:r:\lambda] &= \left[ \frac{v}{k}\frac{v-1}{k-1} : \frac{v-1}{k-1} : 1 \right]\label{eq:SSExactRatio}\\
    &\approx [(e^\epsilon+1)^2 : (e^\epsilon+1) : 1]\label{eq:SSApproxRatio},
\end{align}
when $v \gg 1$. 
Hence, we may want to construct an RPBD scheme satisfying (\ref{eq:SSExactRatio}) to {achieve the exact optimality}. However, we show that such an RPBD scheme must be a block design scheme in the following proposition.
\begin{proposition}\label{prop:SSratioImplyBD}
Let $(\mathcal{X},\mathcal{Y},\mathcal{I})$ be a $(v,b,r,\lambda)$-RPBD. Suppose that there exists an integer $k$ such that
\begin{align}
    [b:r:\lambda] = \left[ \frac{v}{k}\frac{v-1}{k-1} : \frac{v-1}{k-1} : 1 \right].
\end{align}
Then $\mathcal{I}$ is $k$-uniform, hence it is a $(v,b,r,k,\lambda)$-block design.
\end{proposition}
\begin{IEEEproof}
For each $y \in \mathcal{Y}$, let ${k(y)=|\mathcal{I}^y|}$. We need to show that $k(y)=k$ for all $y \in \mathcal{Y}$. The key technique is to show that the empirical variance of $k(y)$ is zero, that is
\begin{align}
    \sigma^2 := \frac{\sum_{y} k(y)^2}{b} - \left(\frac{\sum_{y} k(y)}{b}\right)^2 = 0.
\end{align}
First, by the double counting argument, we can easily observe that
\begin{align}
   |\mathcal{I}| = vr = \sum_{y \in \mathcal{Y}} k(y), \label{eq:doubleCount1}
\end{align}
and
\begin{align}
   |\{(x_1,x_2,y) : (x_1,y) \in \mathcal{I}, (x_2,y) \in \mathcal{I}, x_1 \neq x_2\}|
   =\lambda v(v-1) = \sum_{y \in \mathcal{Y}} k(y)(k(y)-1).
\end{align}
From these, we get
\begin{align}
    \sum_{y \in \mathcal{Y}} k(y)^2 = v(\lambda(v-1)+r),
\end{align}
and
\begin{align}
    \sigma^2 = v \left(\frac{\lambda}{b}(v-1) + \frac{r}{b} - \left(\frac{r}{b}\right)^2 v \right).
\end{align}
By substituting $\frac{\lambda}{b} = \frac{k(k-1)}{v(v-1)}$ and $\frac{r}{b} = \frac{k}{v}$, we get $\sigma^2=0$. Hence, all of $k(y)$ are the same. Also, from (\ref{eq:doubleCount1}), we get $vr=bk(y)$, $k(y)=vr/b = vk/v = k$.    
\end{IEEEproof}

This implies that it is not able to find a non-uniform RPBD scheme marginally equivalent to the optimal SS scheme.
Nevertheless, by the continuity of the worst-case risk \eqref{eq:RPBL2risk_opt}, we expect that an RPBD scheme with the ratio similar to \eqref{eq:SSApproxRatio} {attains near-optimality}.
One easy way to find a near-optimal RPBD scheme with low communication cost is to use an RPBD obtained by truncating a symmetric block design as explained in Remark \ref{remark:truncation}. 
In particular, let $(\mathcal{X}',\mathcal{Y},\mathcal{I}')$ be a $(v',b,r,k,\lambda)$-block design with $v'/k \approx e^\epsilon+1$, and let $(\mathcal{X},\mathcal{Y},\mathcal{I})$ be a $(v,b,r,\lambda)$-RPBD obtained by truncating the block design. Then we have
\begin{align}
    [b:r:\lambda] &= \left[ \frac{v'}{k}\frac{v'-1}{k-1} : \frac{v'-1}{k-1} : 1 \right]\\
    & \approx [(e^\epsilon+1)^2 : (e^\epsilon+1) : 1]
\end{align}
when $v \gg 1$. Hence, if a $(v',b,r,k,\lambda)$-block design gives a block design scheme achieving {near-optimality} for given $\epsilon$, then its truncation gives an RPBD scheme achieving {near-optimality} for each $v<v'$ and the same $\epsilon$. We note that the way that HR and PGR\cite{HR-19Acharya,PGR-22Feldman} are applied to the input data size not of the form of possible input data size for Sylvester's Hadamard matrices or projective geometry-based block design schemes is  equivalent to the use of truncated RPBD schemes. 
{
However, we can also utilize any other block designs in constructing truncated RPBD schemes, and this makes it possible to reduce the communication cost even lower than previously proposed schemes while attaining the {near-optimality}.

For example, suppose we want to find a near-optimal RPBD scheme with low communication cost at $v=100$ and $\epsilon=1$. In this case, PGR\cite{PGR-22Feldman} uses a truncated projective geometry based block design scheme as follows. First, as the block design from projective geometry satisfies $v'/k \approx q$, set $q$ to be the closest prime power to $e^\epsilon+1$, in this case $q=4$. Then, use the smallest $t$ such that $\frac{q^t-1}{q-1} \geq v=100$, that is, $t=5$. Then, truncating such block design becomes an $(100, 341, 85, 21)$-RPBD, and corresponding scheme satisfies $R_{v,2}(Q,\hat{P}_n) = 368.64$. This is only 2\% larger than {the optimal asymptotic risk $M_{v,2,\epsilon} = 360.94$.} Note that the optimal SS for those $v$ and $\epsilon$ uses the communication cost of $b=\binom{100}{27}$, which is translated into $\log_2 b \approx 80.67$ bits per client, while truncated PGR uses the communication cost $b=341$, about 8.41 bits per client.

However, if we employ other block designs, then we can even lower the communication cost while attaining near-optimality.
For the above example, we may instead use an RPBD scheme obtained by truncating a symmetric block design from nonzero quartic residue difference set in Table \ref{tab:blockDesigns} with $v'=101$, which also satisfies $v'/k \approx 4 \approx e^\epsilon+1$. Then, we have a $(100, 101, 25, 6)$-RPBD, and the  corresponding RPBD scheme satisfies lower risk $R_{v,2}(Q,\hat{P}_n) = 362.17$ than truncated PGR, while achieving the communication cost $b=101$, about 6.66 bits per client, which is about 20\% lower number of bits than the truncated PGR.
}

As such, by considering RPBD schemes based on RPBDs obtained by truncating symmetric block designs, {we can construct near-optimal schemes with low communication costs for a larger set of $(v, \epsilon)$.}

\section{Discussions}\label{sec:discussion}
\subsection{Hadamard conjecture and an optimal block design scheme at high privacy regime}\label{subsec:hadConj}
There is a long unsolved conjecture, called the \emph{Hadamard conjecture}\cite{HadamardConjSurvey-04Tressler}. One of the equivalent statements for this conjecture is as follows:
\begin{conjecture}[Hadamard conjecture\cite{HadamardConjSurvey-04Tressler}]
For any positive integer $v \equiv 3 \mod 4$, there is a symmetric block design with $k=(v-1)/2$ and $\lambda=(v-3)/4$.
\end{conjecture}

Remarkably, this conjecture is directly related to the existence of an exactly optimal block design scheme at high privacy regime ($\epsilon \approx 0$) for any input data size $v$ with communication cost at most one bit larger than the input data size. Specifically, the implication of Hadamard conjecture is as follows:
\begin{proposition}
Suppose that the Hadamard conjecture is true. Then, for any $v \geq 2$, the following statements hold: 
\begin{enumerate} 
    \item If $v \equiv 3 \mod 4$ and $\epsilon \leq \frac{1}{2}\log \frac{(v+1)(v+3)}{(v-1)(v-3)}$, there is a symmetric block design scheme which is both exactly optimal and achieving the minimum communication cost of $b=v$.
    \item If $v \equiv 1 \mod 4$ and $\epsilon \leq \frac{1}{2}\log \frac{(v+1)(v+3)}{(v-1)(v-3)}$, there is a block design scheme which is exactly optimal and having the communication cost of $b=2v$.
    \item If $v$ is even and $\epsilon \leq \frac{1}{2}\log \frac{v+2}{v-2}$, there is a block design scheme which is exactly optimal and having the communication cost of $b=2v-2$.
\end{enumerate}

\end{proposition}

\begin{IEEEproof}
First, when $v$ is odd, we construct a block design scheme with $k=(v-1)/2$. Note that for such $k$, we have
\begin{align}
    E(k-1, k;v) &= \sqrt{\frac{(v+1)(v+3)}{(v-1)(v-3)}},\\
    E(k, k+1;v) &= 1.
\end{align}
Thus, by Proposition \ref{prop:optEpsRegion}, such block design scheme is exactly optimal for 
\begin{gather}
 1 \leq e^\epsilon \leq \sqrt{\frac{(v+1)(v+3)}{(v-1)(v-3)}},
\end{gather}
and equivalently,
\begin{equation}\label{eq:odd_opt_reg}
    \epsilon \leq \frac{1}{2}\log \frac{(v+1)(v+3)}{(v-1)(v-3)}.
\end{equation}

Next, when $v$ is even, we construct a block design scheme with $k=v/2$. For such $k$, we have
\begin{align}
    E(k-1, k;v) &= \sqrt{\frac{v+2}{v-2}},\\
    E(k,k+1;v)  &= \sqrt{\frac{v-2}{v+2}} < 1.
\end{align}
Thus, such block design scheme is exactly optimal for 
\begin{gather}
 \sqrt{\frac{v-2}{v+2}} \leq e^\epsilon \leq \sqrt{\frac{v+2}{v-2}},
\end{gather}
and equivalently,
\begin{equation}\label{eq:even_opt_reg}
    \epsilon \leq \frac{1}{2}\log \frac{v+2}{v-2}.
\end{equation}
\begin{enumerate}
    \item If $v \equiv 3 \mod 4$ and $\epsilon \leq \frac{1}{2}\log \frac{(v+1)(v+3)}{(v-1)(v-3)}$, the symmetric block design with $k=(v-1)/2$ from the Hadamard conjecture gives the desired scheme because of \eqref{eq:odd_opt_reg}.
    \item If $v \equiv 1 \mod 4$ and $\epsilon \leq \frac{1}{2}\log \frac{(v+1)(v+3)}{(v-1)(v-3)}$, then $v':=2v+1$ satisfies $v' \equiv 3 \mod 4$. Thus, by the Hadamard conjecture, there is a $(v',b',r',k',\lambda')$-symmetric block design with $k'=(v'-1)/2=v$ and $\lambda' = (v'-3)/4=(v-1)/2$. Then, the derived block design explained in Example \ref{ex:dereived} of such symmetric block design has parameters 
    \begin{align}
        &(v,b,r,k,\lambda) = (k', v'-1, k'-1, \lambda', \lambda'-1)\\
                          &= (v, 2v, v-1, (v-1)/2, (v-3)/2).
    \end{align}
    Combining with \eqref{eq:odd_opt_reg}, such derived block design gives the desired scheme.

    \item If $v$ is even and $\epsilon \leq \frac{1}{2}\log \frac{v+2}{v-2}$, then $v':=2v-1$ satisfies $v' \equiv 3 \mod 4$. Thus, by the Hadamard conjecture, there is a $(v',b',r',k',\lambda')$-symmetric block design with $k'=(v'-1)/2=v-1$ and $\lambda' = (v'-3)/4=(v/2)-1$. Then, the residual block design explained in Example \ref{ex:dereived} of such symmetric block design has parameters 
    \begin{align}
        &(v,b,r,k,\lambda) = (v'-k',v'-1,k',k'-\lambda',\lambda')\\
                          &= (v, 2v-2, v-1, v/2, (v/2)-1).
    \end{align}
    Combining with \eqref{eq:even_opt_reg}, such residual block design gives the desired scheme.
\end{enumerate}
\end{IEEEproof}

\subsection{Other decomposable loss functions}
In this paper, the (near) optimality of block design schemes and RPBD schemes hinges on the {exact} optimality of the SS under the $\ell_u^u$ loss, $1 \leq u \leq 2$  \cite{SSOptim-19MinYe}.
If the SS is {exactly} optimal under some other decomposable loss functions, most of the optimality results in this paper would be extended to such  loss functions. We note that the complete block designs \cite{CombSymmDesigns-06Ionin}, the block designs used for the SS, satisfy more symmetry conditions than general block designs. Due to such a strong symmetry, we conjecture that the SS would also attain the {exact optimality} under other decomposable loss functions with nice symmetry, e.g.,  $\ell_u^u$ loss for $u>2$, although the optimal value of $k$ for given $(v,\epsilon)$ may differ. We provide  two reasons supporting this conjecture.
\begin{itemize}
    \item First, \cite{SSMutualInfoMaximize-16Wang} showed that a SS mechanism with appropriately chosen $k$ maximizes the mutual information between the mechanism input and the output for a uniformly distributed input among all the $\epsilon$-LDP mechanisms, where the appropriate $k$ is in general different from the one achieving the optimal $\ell_2^2$ risks. From the definition of the mutual information, we can show that two block design mechanisms with the same $(v,k,\epsilon)$ also have the same mutual information. Hence, block design mechanisms with appropriate $k$ also achieve the maximum mutual information.
    \item  Second, in the proof of showing the  optimality of PrivUnit$_2$\cite{PrivUnit-18Bhowmick} for mean estimation in \cite{PrivUnitOptim-22Asi}, it was shown that  given a scheme, its `symmetrized' version gives a lower or equal worst-case risk.
    Although this was shown for a different target task of mean estimation and only for the $\ell_2^2$ loss, we expect that the same would hold  for discrete distribution estimation and other decomposable loss functions with nice symmetry. 
\end{itemize}

\section{Conclusion}\label{sec:conclusion}
 In this paper, we presented a new class of LDP schemes for private discrete distribution estimation based on combinatorial block designs, and more generally, regular and  pairwise-balanced designs, which we call block design schemes and RPBD schemes, respectively. In particular, the subset selection (SS) scheme \cite{SS-18MinYe} achieving the {exact optimality} is a special case of the block design schemes and hence also a special case of the RPBD schemes. By showing sufficient and necessary conditions for the marginal equivalence between two block design schemes and for two RPBD schemes, which guarantee that the two schemes achieve  the same risk, we could find many new schemes {which are exactly or nearly} optimal with much lower communication costs than SS.

\bibliographystyle{IEEEtran}
\bibliography{IEEEabrv, ref}

\newpage
\setcounter{page}{1}

\appendices
\section{Proofs}\label{app:proofs}
In this appendix, we give the proofs for Theorems \ref{thm:BDMEstimator}, \ref{thm:BDML2risk}, \ref{thm:blockDesignEquiv}, \ref{thm:RPBL2risk}, and \ref{thm:RPBequiv}. 
As block design schemes are special cases of RPBD schemes, we first prove the theorems for the RPBD schemes, and then extend the proofs to show the theorems for the block design schemes. We first prove Theorems \ref{thm:RPBL2risk} and \ref{thm:RPBequiv}, and using them, we prove Theorems \ref{thm:BDMEstimator}, \ref{thm:BDML2risk}, and \ref{thm:blockDesignEquiv}.

\subsubsection{Proof of Theorem \ref{thm:RPBL2risk}}
By the $r$-regular and $\lambda$-pairwise balanced properties, we have
\begin{align}
Q(\mathcal{I}_x|x) &= \alpha e^\epsilon |\mathcal{I}_x| = \alpha re^\epsilon,
\end{align}
and
\begin{align}
Q(\mathcal{I}_x|x') &= |\mathcal{I}_x \cap \mathcal{I}_{x'}|\alpha e^\epsilon + |\mathcal{I}_x \backslash \mathcal{I}_{x'}|\alpha  \\
&= \alpha(\lambda e^\epsilon + (r-\lambda)),
\end{align}
for all $x \neq x'$.

Therefore, for $X_i \sim P$ and $Y_i \sim Q(\cdot|X_i)$, we have
\begin{IEEEeqnarray}{rl}
&\Pr[(x,Y_i) \in \mathcal{I}] \nonumber\\
=& \sum_{x' \in \mathcal{X}} Q(\mathcal{I}_x | x') \\
=& p_x \alpha re^\epsilon + (1-P_x)\alpha(\lambda e^\epsilon + (r-\lambda))\\
=& \alpha (P_x(r-\lambda)(e^\epsilon-1) + (\lambda e^\epsilon+(r-\lambda))).
\end{IEEEeqnarray}
Thus, $N_x(Y_1,\cdots,Y_n)$ is a binomial random variable with $n$ trials and success probability
\begin{IEEEeqnarray}{c}
    \alpha (P_x(r-\lambda)(e^\epsilon-1) + (\lambda e^\epsilon+(r-\lambda))) .\label{eq:NxSuccessProb}
\end{IEEEeqnarray}
In particular, its mean under $P$ is
\begin{align}
    \mathbb{E}_P[N_x(Y_1,\cdots,Y_n)]
=n\alpha (P_x(r-\lambda)(e^\epsilon-1) + (\lambda e^\epsilon+(r-\lambda))).
\end{align}
Hence, we have
\begin{align}
    P_x = \frac{1}{(r-\lambda)(e^\epsilon-1)}
      \left( \frac{\mathbb{E}_P\left[N_x(Y_1,\cdots,Y_n)\right]}{n\alpha} -(\lambda e^\epsilon+(r-\lambda)) \right).
\end{align}
Thus, the following estimator
\begin{align}\label{eq:Canonical-est-app}
    \hat{P}_{n,x}(Y_1,\cdots,Y_n) = \frac{1}{(r-\lambda)(e^\epsilon-1)}
      \left( \frac{N_x(Y_1,\cdots,Y_n)}{n\alpha} -(\lambda e^\epsilon+(r-\lambda)) \right)
\end{align}
is an unbiased estimator.

Next, let us derive the risk under the mean squared loss. We follow the same procedures in \cite[Theorem 2]{HR-19Acharya}, except that none of upper and lower bounds are taken to derive the risk.
Recall that $N_x(Y_1,\cdots,Y_n)$ is a binomial random variable with $n$ trials and success probability of the form $cP_x + d$, where 
\begin{IEEEeqnarray}{rCl}
    c &=& \alpha (r-\lambda)(e^\epsilon-1),\\
    d &=& \alpha(\lambda e^\epsilon + r-\lambda),\\
    \alpha &=& \frac{1}{re^\epsilon + b - r}.
\end{IEEEeqnarray}
Hence, we get
\begin{IEEEeqnarray}{c}
    \mathbb{E}[N_x(Y_1,\cdots,Y_n)]=n(cP_x+d),\\
    P_x = \frac{\mathbb{E}[N_x(Y_1,\cdots,Y_n)]}{nc}-\frac{d}{c}.
\end{IEEEeqnarray}
Using this, we set the estimator into
\begin{IEEEeqnarray}{c}
    \hat{P}_{n,x}(Y_1,\cdots,Y_n)=\frac{N_x(Y_1,\cdots,Y_n)}{nc}-\frac{d}{c},
\end{IEEEeqnarray}
so that it is unbiased.

Since this estimator is unbiased, the risk under the $\ell_2^2$ loss is the same as the sum of the variances of all $\hat{P}_x(Y_1,\cdots,Y_n)$. Therefore, we have
\begin{IEEEeqnarray}{rl}
    &R_{v, 2}^n(P, Q, \hat{P}_n) \nonumber\\
    =&\sum_{x \in \mathcal{X}}\mathrm{Var}[\hat{P}_{n,x}(Y_1,\cdots,Y_n)]\\
    =&\frac{1}{n^2 c^2} \sum_{x \in \mathcal{X}}\mathrm{Var}[N_x(Y_1,\cdots,Y_n)]\\
    =&\frac{1}{n^2 c^2}\sum_{x \in \mathcal{X}}n(cP_x+d)(1-(cP_x+d))\\
    =&\frac{1}{n c^2}\sum_{x \in \mathcal{X}}(cP_x+d)(1-(cP_x+d))\label{eq:L2CalcIntermediate}\\
    =&\frac{1}{n c^2}\sum_{x \in \mathcal{X}}(-c^2 P_x^2 + \gamma_1 P_x + \gamma_0)\\
    =&\frac{1}{n c^2}\left(-c^2 \sum_{x \in \mathcal{X}}P_x^2 + \gamma_1+ \gamma_0 v\right)\\
    =&\frac{1}{n}\left(\delta - \sum_{x \in \mathcal{X}}P_x^2\right), \label{eq:L2riskWithConstant}
\end{IEEEeqnarray}
where $\gamma_1,\gamma_0,\delta$ are constants which depend on parameters and do not depend on $P$.
Hence, the worst-case risk is attained when $\sum_{x \in \mathcal{X}}P_x^2$ is minimized. Since $\sum_{x \in \mathcal{X}}P_x = 1$, by the Cauchy-Schwartz inequality, we have
\begin{IEEEeqnarray}{c}
    \sum_{x \in \mathcal{X}}P_x^2 \geq \frac{1}{\left(\sum_{x \in \mathcal{X}}1\right)}=\frac{1}{v},
\end{IEEEeqnarray}
and the equality is attained when $P_x=\frac{1}{v}$ for all $x \in \mathcal{X}$. Thus, the worst-case risk is attained when $P$ is the uniform distribution.

From equation (\ref{eq:L2CalcIntermediate}) and substituting $P_x=1/v$, we have
\begin{IEEEeqnarray}{rl}
    &R_{v, 2}^n(Q, \hat{P}_n) \nonumber\\
    =&\frac{1}{n c^2}\sum_{x \in \mathcal{X}}\left(\frac{c}{v}+d\right)\left(1-\left(\frac{c}{v}+d\right)\right)\\
    =&\frac{v}{n c^2}\left(\frac{c}{v}+d\right)\left(1-\left(\frac{c}{v}+d\right)\right)\\
    =&\frac{1}{nc^2 v}(c+dv)(v-c-dv).
\end{IEEEeqnarray}
Observe that $c+d=\alpha re^\epsilon$. Using this, we have
\begin{IEEEeqnarray}{rl}
    &c+dv \nonumber\\
    =& (c+d)+(v-1)d\\
    =& \alpha re^\epsilon + (v-1)\alpha(\lambda e^\epsilon+r-\lambda)\\
    =& \alpha(re^\epsilon+(v-1)(\lambda e^\epsilon+r-\lambda)),
\end{IEEEeqnarray}
and
\begin{IEEEeqnarray}{rl}
    &v-c-dv \nonumber\\
    =& v - \alpha(re^\epsilon+(v-1)(\lambda e^\epsilon+r-\lambda))\\
    =& \alpha\left(\frac{v}{\alpha}-(re^\epsilon+(v-1)(\lambda e^\epsilon+r-\lambda)) \right)\\
    =& \alpha\left(\frac{1}{\alpha}-re^\epsilon - (v-1)\left(\lambda e^\epsilon+r-\lambda-\frac{1}{\alpha}\right) \right)\\
    =& \alpha(b-r - (v-1)\left((\lambda-r)(e^\epsilon-1)-(b-r) \right))\\
    =& \alpha(v(b-r)-(v-1)(\lambda-r)(e^\epsilon-1)).
\end{IEEEeqnarray}
Hence, we have
    \begin{gather}
        R_{v, 2}^n(Q, \hat{P}_n) = \frac{1}{nc^2 v}(c+dv)(v-c-dv)\\
        = \frac{re^\epsilon+(v-1)(\lambda e^\epsilon+r-\lambda)}{(r-\lambda)^2(e^\epsilon-1)^2 nk} \left[v(b-r) + (v-1)(r-\lambda)(e^\epsilon-1) \right]. \nonumber
    \end{gather}
Also, from equation (\ref{eq:L2riskWithConstant}), we have
\begin{IEEEeqnarray}{rCl}
    R_{v, 2}^n(Q, \hat{P}) &=& R_{v, 2}^n(P, Q, \hat{P})|_{P_x=1/v}\\
    &=&\frac{1}{n}\left(\delta-\sum_{x \in \mathcal{X}}\frac{1}{v^2}\right)\\
    &=&\frac{1}{n}\left(\delta-\frac{v}{v^2}\right)\\
    &=&\frac{1}{n}\left(\delta-\frac{1}{v}\right).
\end{IEEEeqnarray}
Thus, we have
\begin{align}
\delta = \frac{re^\epsilon+(v-1)(\lambda e^\epsilon+r-\lambda)}{(r-\lambda)^2(e^\epsilon-1)^2 v}\left[v(b-r) + (v-1)(r-\lambda)(e^\epsilon-1) \right]+ \frac{1}{v}.
\end{align}
This proves the theorem. $\IEEEQEDhere$

\subsubsection{Proof of Theorem \ref{thm:RPBequiv}}
Let $(Q,\hat{P}_n)$ and $(Q',\hat{P}_n')$ be RPBD schemes with parameters $(v, b, r, \lambda, \epsilon)$ and $(v, b',r',\lambda', \epsilon)$, constructed from incidence structures $\mathcal{I},\mathcal{I}'$, respectively.

Let ${\alpha=1/(re^\epsilon+b-r)}$ and ${\alpha'=1/(r'e^\epsilon+b'-r')}$ be the normalization constants for $Q$ and $Q'$, respectively. Also, let 
\begin{align}
    N_x(Y_1,\cdots,Y_n) &= \sum_{i=1}^{n}\mathbbm{1}(Y_i \in \mathcal{I}_x),\\
    N_x'(Y_1',\cdots,Y_n') &= \sum_{i=1}^{n}\mathbbm{1}(Y_i' \in \mathcal{I}'_x).
\end{align}
First, assume that $(b',r',\lambda')=t(b,r,\lambda)$, $t>0$.
By the assumption, we have $\alpha'=\frac{1}{t}\alpha$. By comparing the success probability in (\ref{eq:NxSuccessProb}), we can see that
\begin{align}
\alpha (P_x(r-\lambda)(e^\epsilon-1) + (\lambda e^\epsilon+(r-\lambda)))
=\alpha' (P_x(r'-\lambda')(e^\epsilon-1) + (\lambda' e^\epsilon+(r'-\lambda'))),
\end{align}
for all $P_x \in [0,1]$. Hence
\begin{align}
N_x(Y_1,\cdots,Y_n) \overset{d}{=} N_x'(Y_1',\cdots,Y_n'), \label{eq:NxEquiv}
\end{align}
for all $P \in \Delta_v$.

Next, let us compare the formula of the estimator in (\ref{eq:Canonical-est-app}). The formula is of the form
\begin{align}
    \hat{P}_{n,x} = g \frac{N_x}{n} - h, \quad \hat{P}'_{n,x} = g'\frac{N_x'}{n} - h', \label{eq:estFormulaSimple}
\end{align}
where
\begin{align}
    g &= \frac{1}{\alpha} \frac{1}{(r-\lambda)(e^\epsilon-1)}, &h&=\frac{\lambda e^\epsilon + (r-\lambda)}{(r-\lambda)(e^\epsilon-1)},\label{eq:estFormulaCoeffs1}\\
    g' &= \frac{1}{\alpha'} \frac{1}{(r'-\lambda')(e^\epsilon-1)}, &h'&=\frac{\lambda' e^\epsilon + (r'-\lambda')}{(r'-\lambda')(e^\epsilon-1)}.\label{eq:estFormulaCoeffs2}
\end{align}
Again, we can see that $g=g'$ and $h=h'$. Together with (\ref{eq:NxEquiv}), we get the desired marginal equivalence. 

Conversely, assume that $(Q,\hat{P}_n)$ and $(Q',\hat{P}_n')$ are marginally equivalent. From $b>r>\lambda$, we can show that \eqref{eq:NxSuccessProb} is neither $0$ nor $1$ for $P_x=0$. Thus, $N_x(Y_1,\cdots,Y_n)$ have support $\{0,1,\cdots,n\}$. From (\ref{eq:estFormulaSimple}), the estimator $\hat{P}_{n,x}(Y_1,\cdots,Y_n)$ have support $\{-h,-h+g/n,-h+2g/n,\cdots, g-h\}$. Similarly, $\hat{P}_{n,x}'(Y_1',\cdots,Y_n')$ have support $\{-h',-h'+g'/n,-h'+2g'/n,\cdots, g'-h'\}$. For the marginal equivalence, the support must be the same. Thus, either
\begin{align}
    -h=-h',\quad g-h=g'-h', \label{eq:sameSupportCond1}
\end{align}
or
\begin{align}
    -h=g'-h',\quad g-h=-h'. \label{eq:sameSupportCond2}
\end{align}
Note that from $b>r>\lambda$, we have $\alpha,g,h,g',h' > 0$. Hence, (\ref{eq:sameSupportCond2}) implies $g=h-h'=-g'$, which is impossible. Thus we must have (\ref{eq:sameSupportCond1}). This implies $g=g'$ and $h=h'$.

Now, by fixing $b',r',\lambda',g',h'$, the equations 
\begin{align}
    g = \frac{re^\epsilon+b-r}{(r-\lambda)(e^\epsilon-1)} = g'\\
    h = \frac{\lambda e^\epsilon + (r-\lambda)}{(r-\lambda)(e^\epsilon-1)} = h'
\end{align}
can be seen as a system of linear equations in $(b,r,\lambda)$:
\begin{equation}
    \begin{cases}b + (e^\epsilon-1)(1-g')r + (e^\epsilon-1)g' \lambda = 0\\
      (1-(e^\epsilon-1)h')r   + (e^\epsilon-1)(1+h')\lambda = 0 \end{cases},
\end{equation}
which has a particular solution $(b,r,\lambda)=(b',r',\lambda')$. Since $(e^\epsilon-1)(1+h')>0$, it can be easily observed that the matrix of the coefficients for this system has rank $2$. Hence, the dimension of the solution space is $1$. Thus, $(b,r,\lambda)$ must be a constant multiple of a particular solution $(b',r',\lambda')$. Since $b,b'>0$, the constant multiplied must be a positive number. This ends the proof of the theorem. $\IEEEQEDhere$

\subsubsection{Proof of Theorems \ref{thm:BDMEstimator}, \ref{thm:BDML2risk} and \ref{thm:blockDesignEquiv}}
First, let us prove Theorem \ref{thm:blockDesignEquiv}. From Lemma \ref{lem:blockDesignFundEq}, it can be easily checked that
\begin{IEEEeqnarray}{c}
    (b,r,\lambda) = t(v(v-1), k(v-1), k(k-1)),\\
    (b',r',\lambda')=t'(v(v-1),k'(v-1),k'(k'-1)),
\end{IEEEeqnarray}
for $t=\frac{b}{v(v-1)}$, $t'=\frac{b'}{v(v-1)}$.
Hence, the desired necessary and sufficient condition for marginal equivalence directly follows from Theorem \ref{thm:RPBequiv}.

For Theorem \ref{thm:BDMEstimator}, the formula of the unbiased estimator is the same as Theorem \ref{thm:RPBL2risk}. For Theorem \ref{thm:BDML2risk}, the formula of the risk is derived by substitution of
\begin{align}
(b,r,\lambda) = t(v(v-1), k(v-1), k(k-1))
\end{align}
into the formula in Theorem \ref{thm:RPBL2risk} and cancellation of $t$. $\IEEEQEDhere$

\section{Descriptions of block designs}\label{app:blockDesignsDescription}
The proofs showing that the below incidence structures indeed satisfy the properties of block designs can be found in \cite{DifferenceSets-13Moore, CombSymmDesigns-06Ionin}.
\subsubsection{Trivial symmetric block design}
Let $v \geq 2$ be given.
For $\mathcal{X}=\{1,2,\cdots,v\}$, let $\mathcal{Y}=\mathcal{X}$, and $\mathcal{I} = \{(x,y) : x=y\}.$
This is a symmetric block design with $k=1$.

\subsubsection{Complete block design}

This is in general non-symmetric block design. Let $1 \leq k < v$ be given. For $\mathcal{X}=\{1,2,\cdots,v\}$, let $\mathcal{Y}$ be the collection of all subsets of $\mathcal{X}$ of size $k$, that is
\begin{IEEEeqnarray}{c}
    \mathcal{Y} = \binom{\mathcal{X}}{k}.
\end{IEEEeqnarray}
Each $y \in \mathcal{Y}$ is a subset of $\mathcal{X}$.
An incidence structure is defined by
\begin{IEEEeqnarray}{c}
    \mathcal{I} = \{(x,y) : x \in y\}.
\end{IEEEeqnarray}
Note that the complete block design with $k=1$ is the same as the trivial symmetric block design.

\subsubsection{Projective geometry}
Let $q,t$ be positive integers such that $q$ is a prime power and $t \geq 2$.
Let $\mathbb{F}_q$ be the finite field of order $q$, and let $P(\mathbb{F}_q^t)$ be the set of all 1-dimensional subspaces of $t$-dimensional vector space $\mathbb{F}_q^t$, called the projective space. We identify $\mathcal{X}$ as this projective space $\mathcal{X}=P(\mathbb{F}_q^t)$, which has size $\frac{q^t-1}{q-1}$. Also, let $\mathcal{Y}$ be the collection of all $(t-1)$-dimensional subspaces of $\mathbb{F}_q^t$. By letting
\begin{IEEEeqnarray}{c}
    \mathcal{I} = \{(x,y) : x \subset y\},
\end{IEEEeqnarray}
this is a symmetric block design with $v=\frac{q^t-1}{q-1}$, $k=\frac{q^{t-1}-1}{q-1}$.

\subsubsection{Sylvester's Hadamard matrix}

Define the Sylvester's Hadamard matrix as follows:
Let $H_2=\begin{pmatrix} 1 & 1 \\ 1 & -1 \end{pmatrix}$, and for each $t \geq 2$, recursively define
\begin{IEEEeqnarray}{c}
    H_{2^t} = \begin{pmatrix} H_{2^{t-1}} & H_{2^{t-1}} \\ H_{2^{t-1}} & -H_{2^{t-1}} \end{pmatrix}.
\end{IEEEeqnarray}

Let $t \geq 2$. We identify $\mathcal{X}=\mathcal{Y}=\{2,3,4,\cdots,2^t\}$, and define 
\begin{align}
    \mathcal{I}=  \{(x,y):\text{the } (x,y)\text{'th component of }H_{2^t}\text{ is }1\}.
\end{align}
This is a symmetric block design with $v=2^t-1$, $k=2^{t-1}-1$.
\subsubsection{Paley's Hadamard matrix}
Although the description by Hadamard matrix is well-known, we present an alternative but equivalent description by difference set\cite{DifferenceSets-13Moore}, which gives an easy way to implement the scheme.

Let $v$ be a prime power such that $v \equiv 3 \mod 4$. We identify $\mathcal{X}=\mathcal{Y}=\mathbb{F}_v$, the finite field of order $v$.
By letting 
\begin{IEEEeqnarray}{c}
    \mathcal{I}=\{(x,y):y-x=a^2\text{ for some }a \in \mathbb{F}_v^\times\},
\end{IEEEeqnarray}
this is a symmetric block design with $k=(v-1)/2$.

\subsubsection{Twin prime power difference set}

Let $q$ be an odd natural number such that both $q$ and $q+2$ are prime powers. 
We identify $\mathcal{X}=\mathcal{Y}=\mathbb{F}_q \times \mathbb{F}_{q+2}$, the direct product of finite fields of order $q$ and $q+2$.
Define $\mathcal{I}$ to be the set of all $(x,y)$ such that
\begin{IEEEeqnarray}{c}
    y-x=(a_1,a_2) \in \mathbb{F}_q \times \mathbb{F}_{q+2}
\end{IEEEeqnarray}
implies $(a_1,a_2)$ to satisfy either one of the following three conditions:
\begin{itemize}
    \item $a_2=0$.
    \item both $a_1,a_2$ are non-zero squares in $\mathbb{F}_q^\times,\mathbb{F}_{q+2}^\times$ respectively.
    \item both $a_1,a_2$ are non-zero non-squares in $\mathbb{F}_q^\times,\mathbb{F}_{q+2}^\times$ respectively.
\end{itemize}
This is a symmetric block design with $v=q(q+2)$, ${k=\frac{q(q+2)-1}{2}}$.

\subsubsection{Nonzero quartic residue difference set}

Let $v$ be a prime power such that $v=4t^2+1$ for some odd integer $t$. We identify $\mathcal{X}=\mathcal{Y}=\mathbb{F}_v$, the finite field of order $v$. Define 
\begin{IEEEeqnarray}{c}
    \mathcal{I}=\{(x,y):y-x=a^4\text{ for some }a \in \mathbb{F}_v^\times\}.
\end{IEEEeqnarray}    
This is a symmetric block design with $k=(v-1)/4$.

\subsubsection{(Including zero) Quartic residue difference set}

Let $v$ be a prime power such that $v=4t^2+9$ for some odd integer $t$. We identify $\mathcal{X}=\mathcal{Y}=\mathbb{F}_v$, the finite field of order $v$. We define
\begin{IEEEeqnarray}{c}
    \mathcal{I}=\{(x,y):y-x=a^4\text{ for some }a \in \mathbb{F}_v\}.
\end{IEEEeqnarray}
Note that we allow $a=0$ in this definition. This is a symmetric block design with $k=(v+3)/4$.

\section{Algorithms for newly proposed block design schemes}\label{app:blockDesignsAlgorithms}
{
In this appendix, we will present algorithms for block designs schemes arise from Paley's Hadamard matrix, Twin prime power difference set, and quartic residue difference sets of two versions described in Appendix \ref{app:blockDesignsDescription}.
Also, as in \cite{HR-19Acharya}, \cite{PGR-22Feldman}, we will briefly analyze the computational complexity of the server algorithm to perform estimation, although we also present a client algorithm for the perturbation. In our list of block designs, finite fields of order at most $v$ are involved. As in \cite{PGR-22Feldman}, we assume that we can perform finite field arithmetics in $\mathbb{F}_q$ in $O(1)$ time after preprocessing, by finding a primitive element $g$ of $\mathbb{F}_q$ and generate a lookup table for the exponent of each element of $\mathbb{F}_q^\times$ (here, the primitive element is the generator of the cyclic group $\mathbb{F}_q^\times$, and the exponent of an element $a \in \mathbb{F}_q^\times$ with respect to a primitive element $g$ is an integer $m \in \{0,1,\cdots,q-2\}$ such that $g^m = a$).

Aforementioned block designs are all based on a special mathematical structure called the \emph{difference set}\cite{DifferenceSets-13Moore}, described as follows:
\begin{enumerate}
    \item $\mathcal{X}=\mathcal{Y}$, and it has an additive group structure.
    \item $\mathcal{I} = \{(x,y) : y-x \in \mathcal{D}\}$ for some $\mathcal{D} \subset \mathcal{X}$.
\end{enumerate}
The set $\mathcal{D}$ for each block designs are
\begin{enumerate}
    \item Paley's Hadamard matrix based block design

    $\mathcal{D} = \{a^2 : a \in \mathbb{F}_v^\times\}$

    \item Twin prime power difference set based block design
    
    $\mathcal{D} = \{(a_1,0) : a_1 \in \mathbb{F}_q^\times\} \cup \{(a_1,a_2) : a_1 \in (\mathbb{F}_q^\times)^2, a_2 \in (\mathbb{F}_{q+2}^\times)^2\} \cup \{(a_1,a_2) : a_1 \in \mathbb{F}_q^\times \backslash (\mathbb{F}_q^\times)^2, a_2 \in \mathbb{F}_{q+2}^\times \backslash (\mathbb{F}_{q+2}^\times)^2\}$

    which can be alternatively written as

    $\mathcal{D} = \{(a_1,0) : a_1 \in \mathbb{F}_q^\times\} \cup \{(a_1^2,a_2^2) : a_1 \in \mathbb{F}_q^\times, a_2 \in \mathbb{F}_{q+2}^\times\} \cup \{(a_1^2 g_1,a_2^2 g_2) : a_1 \in \mathbb{F}_q^\times, a_2 \in \mathbb{F}_{q+2}^\times\}$

    where $g_1,g_2$ are primitive elements in $\mathbb{F}_q,\mathbb{F}_{q+2}$, respectively.

    \item Nonzero quartic residue difference set based block design
    
    $\mathcal{D} = \{a^4 : a \in \mathbb{F}_v^\times\}$

    \item (Including zero) quartic residue difference set based block design
    
    $\mathcal{D} = \{a^4 : a \in \mathbb{F}_v\}$
\end{enumerate}

If a block design has a difference set structure as above, then we can implement the corresponding block design scheme as follows
\begin{enumerate}
    \item Perturbation algorithm for client

    For given $X$, we can draw $Y \sim Q(\cdot|X)$ as follows:
    \begin{itemize}
        
        \item First, draw an independent Bernoulli random variable $B$ with $\mathrm{Pr}(B=1) = \frac{re^\epsilon-r}{re^\epsilon+b-r}$.
        \item If $B=0$, then draw $Y$ uniformly random from $\mathcal{Y}$, independent of $X$.

        If $B=1$, then first draw $D$ uniformly random from $\mathcal{D}$, independent of $X$. Then, let $Y=X+D$.
    \end{itemize}
    
    \item Estimation algorithm for server

    Given $(Y_1,Y_2,\cdots,Y_n)$, we first count the occurrence of each $y \in \mathcal{Y}$, producing $C_y=\sum_{i=1}^n \mathbbm{1}(Y_i=y)$ for each $y \in \mathcal{Y}$. This is a standard first step performed also in \cite{HR-19Acharya, PGR-22Feldman}, running in $O(n)$ times. After that, we calculate $N_x (Y_1,\cdots,Y_n) = \sum_{y:(x,y) \in \mathcal{I}} C_y$ for each $x$. By our construction of $\mathcal{I}$, we have
    \begin{align}
        N_x = \sum_{d \in \mathcal{D}} C_{y-d}. \label{eq:NxFormula}
    \end{align}
    After that, we calculate each $\hat{P}_x$ from $N_x$ by the linear function as in \eqref{eq:Canonical est}, which runs in $O(v)$ times.
    
\end{enumerate}

For the client side, the uniform sampling of an element $D$ from $\mathcal{D}$ can be performed by the following algorithms:

\begin{enumerate}
    \item Paley's Hadamard matrix based block design

    \begin{itemize}
        \item First, draw $E$ uniformly random from $\mathbb{F}_v^\times$.
        \item Then, set $D=E^2$.
    \end{itemize}

    \item Twin prime power difference set based block design
    \begin{itemize}
        \item Draw two independent Bernoulli random variables $B_1,B_2$, where $\mathrm{Pr}(B_1=1)=\frac{q}{r}$ and $\mathrm{Pr}(B_2=1)=\frac{1}{2}$.
        \item If $B_1=1$, then draw $D_1$ uniformly random from $\mathbb{F}_q$, and set $D=(D_1,0)$.
        \item If $B_1=0$, then draw $E_1,E_2$ independently and uniform randomly from $\mathbb{F}_q^\times, \mathbb{F}_{q+2}^\times$, respectively. After that
        \begin{itemize}
            \item If $B_2=0$, then set $D=(E_1^2,E_2^2)$.
            \item If $B_2=1$, then set $D=(E_1^2 g_1, E_2^2 g_2)$, where $g_1,g_2$ are primitive elements of $\mathbb{F}_q$ and $\mathbb{F}_{q+2}$, respectively.
        \end{itemize}
    \end{itemize}

    \item Nonzero quartic residue difference set based block design
    \begin{itemize}
        \item First, draw $E$ uniformly random from $\mathbb{F}_v^\times$.
        \item Then, set $D=E^4$.
    \end{itemize}

    \item (Including zero) quartic residue difference set based block design
    \begin{itemize}
        \item Draw an independent Bernoulli random variable $B_1$ with $\mathrm{Pr}(B_1=1)=\frac{1}{r}$.
        \item If $B_1=1$, then set $D=0$.
        \item If $B_1=0$, then draw $E$ uniformly random from $\mathbb{F}_v^\times$, and set $D=E^4$.
    \end{itemize}
\end{enumerate}
For the server side, it remains to present an algorithm to calculate \eqref{eq:NxFormula}. Since the additive group of a finite field is a direct product of cyclic groups (of prime orders), the additive group structure of $\mathcal{X}$ in block designs in our interests are also direct products of cyclic groups.
Thus, the expression \eqref{eq:NxFormula} is nothing but the multidimensional circular convolution of $\{C_y\}_y$ and $\{D_y\}_y$ such that $D_y=1$ if $y \in \mathcal{D}$, and $D_y=0$ otherwise.
Hence, once we construct $\{D_y\}_y$, we can calculate $\{N_x\}_x$ as in \eqref{eq:NxFormula} in $O(v\log v)$ times by using the standard convolution theorem and the fast Fourier transform.
We can construct $\{D_y\}_y$ by an enumeration of elements of $\mathcal{D}$, which can be done in $O(v)$ times by using definitions of $\mathcal{D}$ for each block designs.
In total, server uses $O(n+v\log v)$ times to perform an estimation, which is comparable to previously proposed schemes\cite{HR-19Acharya, PGR-22Feldman}.

\begin{remark}
    The projective geometry based block design can be also equipped with a difference set structure, called the Singer's difference set\cite{DifferenceSets-13Moore}. By using this, we can present another way to implement PGR\cite{PGR-22Feldman} efficiently by the similar way as above.
\end{remark}
}

\end{document}